\documentclass[twocolumn, twocolappendix]{aastex631}

\received{2022 October 26}
\accepted{2022 December 13}

\shorttitle{\texttt{NICpolpy}: NIC Polarimetry Reducer}
\shortauthors{Y. P. Bach et al.}
\graphicspath{{./}{figures/}}

\begin{document}

\title{\large Data Reduction Process and Pipeline for the NIC Polarimetry Mode in Python, \texttt{NICpolpy}}

\author[0000-0002-2618-1124]{{\rm Yoonsoo} P. Bach}
\affiliation{Astronomy Program, Department of Physics and Astronomy, Seoul National University, 
  \\Gwanak-ro 1, Gwanak-gu, Seoul 08826, Republic of Korea}
\affiliation{SNU Astronomy Research Center, Department of Physics and Astronomy, Seoul National University,
  \\Gwanak-ro 1, Gwanak-gu, Seoul 08826, Republic of Korea}
\email{ysbach93@gmail.com, ishiguro@snu.ac.kr, takahashi@nhao.jp}

\author[0000-0002-7332-2479]{{\rm Masateru} Ishiguro}
\affiliation{Astronomy Program, Department of Physics and Astronomy, Seoul National University, 
  \\Gwanak-ro 1, Gwanak-gu, Seoul 08826, Republic of Korea}
\affiliation{SNU Astronomy Research Center, Department of Physics and Astronomy, Seoul National University,
  \\Gwanak-ro 1, Gwanak-gu, Seoul 08826, Republic of Korea}

\author[0000-0002-2928-8306]{{\rm Jun} Takahashi}
\affiliation{Nishi-Harima Astronomical Observatory, Center for Astronomy, University of Hyogo,\\
                   407--2 Nishigaichi, Sayo-cho, Hyogo 679--5313, Japan}

\author[0000-0002-3291-4056]{{\rm Jooyeon} Geem}
\affiliation{Astronomy Program, Department of Physics and Astronomy, Seoul National University, 
  \\Gwanak-ro 1, Gwanak-gu, Seoul 08826, Republic of Korea}
\affiliation{SNU Astronomy Research Center, Department of Physics and Astronomy, Seoul National University,
  \\Gwanak-ro 1, Gwanak-gu, Seoul 08826, Republic of Korea}



\begin{abstract}
A systematic way of data reduction for the NIC polarimetry mode has been devised and implemented to an open software called \texttt{NICpolpy} in the programming language python (tested on version 3.8--3.10 as of writing). On top of the classical methods, including vertical pattern removal, a new way of diagonal pattern (Fourier pattern) removal has been implemented. Each image undergoes four reduction steps, resulting in ``level 1'' to ``level 4'' products, as well as nightly calibration frames. A simple tutorial and in-depth descriptions are provided, as well as the descriptions of algorithms. The dome flat frames (taken on UT 2020-06-03) were analyzed, and the pixel positions vulnerable to flat error were found. Using the dark and flat frames, the detector parameters, gain factor (the conversion factor), and readout noise are also updated. We found gain factor and readout noise are likely constants over pixel or ``quadrant''.
\end{abstract}



\section{Introduction}
In this work, we describe a way to reduce Nishi-Harima Astronomical Observatory (NHAO) Nishiharima Infrared Camera (NIC) polarimetric mode data. The process is implemented to \texttt{NICpolpy}\footnote{
    The stable version is registered to the Python Package Index (PyPI): \texttt{https://pypi.org/project/NICpolpy/}. The development version is available via GitHub at \texttt{https://github.com/ysBach/NICpolpy}. See Sect. \ref{s: code usage} for details.
}, and the implementation details are given. The term \textit{reduction} in this work is restricted to preprocessing of image data, including artifact removal and the standard dark and flat corrections, but excluding photometry/polarimetry and error analysis. We analyzed the dome flat frames (UT 2020-06-03) and examined if any pixels have large uncertainty due to imperfect flat correction or hot pixels. Also, the effect of the instrumental rotator and half-wave plate (HWP) angles are shown. 

During the development of \texttt{NICpolpy}, the detector parameters, viz., gain factor (conversion factor, unit of electrons per ADU) and readout noise (unit of electrons), were recalculated using dark frames (UT 2019-11-21) and the dome flat frames. 
These had already been determined in \citet{Ishiguro2011ARNHAO} more than a decade ago and should be updated due to a possible secular change. Throughout this work, the three detectors associated with three filters, $ \mathrm{J} $, $ \mathrm{H} $, and $ \mathrm{K_s} $, are referred to as J-, H-, and K-band, respectively.

In Sect. \ref{s: dataprep}, the raw image and reduction processes implemented to \texttt{NICpolpy} package are described. Then the practical usage of the package is shown in Sect. \ref{s: code usage}, with additional descriptions of algorithms and implementations. Sect. \ref{s: flat} describes an analysis of polarimetric dome flat images to find any locations vulnerable to flat fielding uncertainties. Finally, in Sect. \ref{s: gain rdn}, a new estimation of instrument parameters (gain and readout noise), as well as the corresponding statistical analyses, are described. 

\section{Description of Image and Data Reduction}\label{s: dataprep}
In \texttt{NICpolpy}, the original image is called \textit{level 0} or \texttt{lv0}. Unfortunately, level 0 images are saved in 32-bit integer format (\texttt{BITPIX=32}), unnecessarily doubling the file size (a file is 4.2 MB instead of 2.1 MB). This is because NIC does not produce bias frames, so the pixel values can be negative, while the standard FITS format does not support a signed 16-bit integer. One possibility is to add an arbitrary constant bias (e.g., few thousands) and save as an unsigned 16-bit integer ($ 0 \mathrm{-} 32,767 $), because anyway the maximum signal for NIC is $ < 2 \times 10^4\,\mathrm{ADU} $ \citep{Ishiguro2011ARNHAO}. Another possibility is to use \texttt{BZERO} key in the FITS header. At this moment, neither is implemented to \texttt{NICpolpy} because it is something to be implemented when the data are saved in the first place.

After the column-pattern subtraction, it is now called \textit{level 1} or \texttt{lv1}. The columnar pattern is called \textit{vertical pattern} in \texttt{NICpolpy}. Then an additional semi-periodic noise, the so-called \textit{Fourier pattern}, across the detector frame remains. This is completely random for each frame (although the exact cause is unsolved) and has an amplitude of a few analog-to-digital units (ADU). After this pattern is removed, the image is called \textit{level 2} or \texttt{lv2}. Then, usual dark subtraction, flat fielding, and bad pixel interpolation results in \textit{level 3} or \texttt{lv3}. Finally, after an optional fringe subtraction and cosmic-ray removal, the result is called \textit{level 4} or \texttt{lv4}. 

\begin{figure*}
  \centering
  \includegraphics[width=\linewidth]{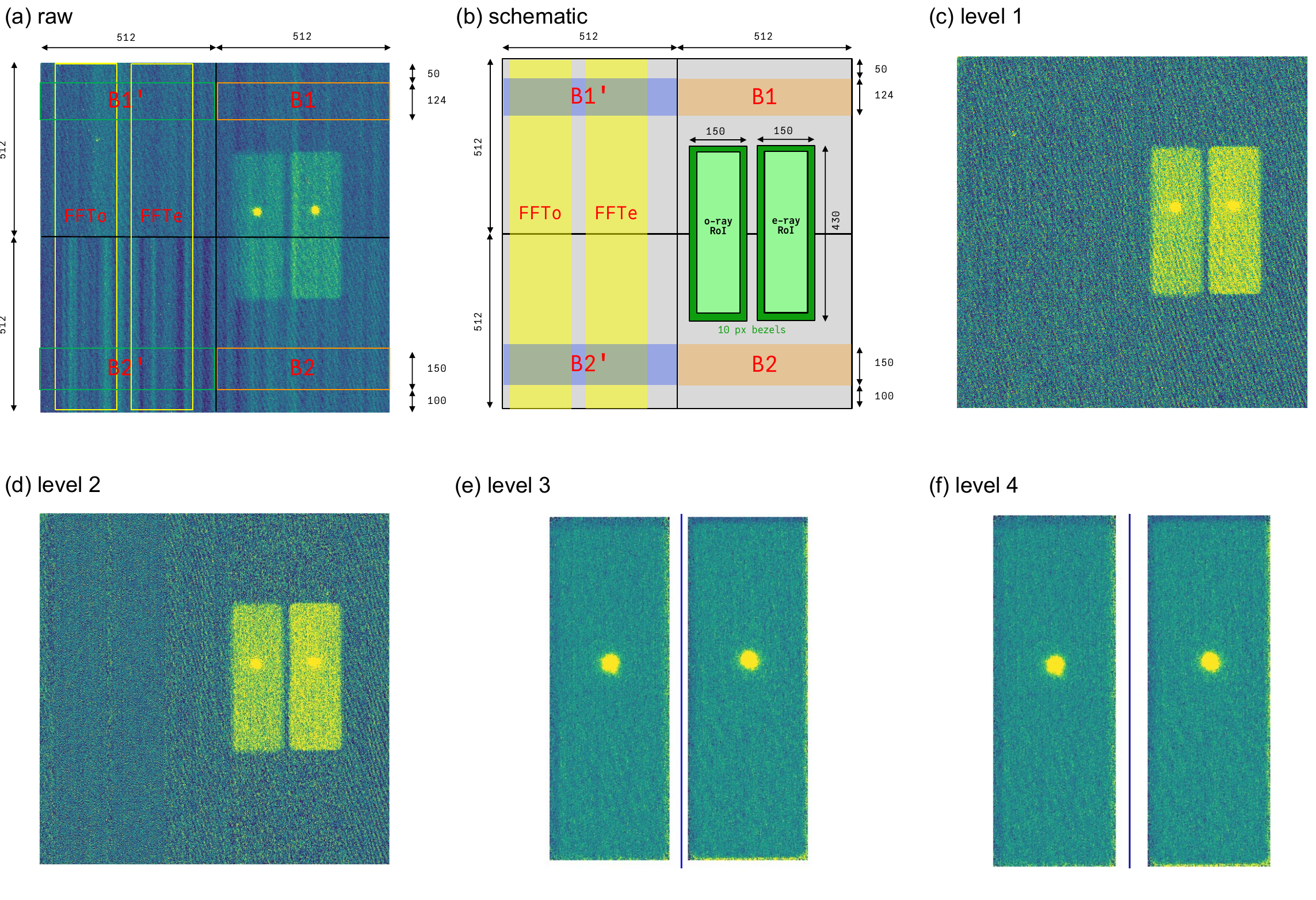}
  \caption{An example exposure (2019-10-22 UT, Vesta). (a) The raw data, (b) schematic diagram describing the full NIC frame in polarimetry mode, (c) level 1, (d) level 2, (e) level 3, and (f) level 4. All images are in the same limits and scale. Numbers in (a) and (b) are in the unit of a pixel, showing the dimensions of indicated areas. Light comes only to some part on the right half of the full array, as in (a) called the lit part. These are the green-boxed regions in (b). The regions indicated by letter B in (a) and (b) are used to infer the vertical pattern, resulting in (c). The left side of the array, indicated by FFTo and FFTe in (a) and (b), is used to infer the Fourier pattern, resulting in (d). Figures (c) and (d) are scaled identically such that the Fourier pattern along the diagonal direction is exaggerated. Note that the pattern is reduced in the FFTo/FFTe areas and lit parts in (d), compared to (c). After splitting o-/e-ray regions, standard flat fielding, and bad-pixel interpolation, level 3 is obtained (e). Finally, a conventional cosmic-ray rejection technique is applied to obtain level 4 (f). In this example, fringe removal was turned off and no cosmic ray was detected, so the images in (e) and (f) are identical. The dark green areas in (b) indicate the 10-pixel bezels that are ignored (cropped) in any data analyses after level 4. See text for details.}
  \label{fig:fignicfitsanatomy}
\end{figure*}

Fig. \ref{fig:fignicfitsanatomy} shows a sample FITS frame during the reduction steps. The raw frame consist of four parts of size $ 512 \times 512 \,\mathrm{pix}$, and we call them \textit{quadrants}. As is visible, the light from the exposure falls onto a small portion in the right half of the full array, indicated as green boxes in Fig. \ref{fig:fignicfitsanatomy}b. The dark green part is called the \textit{lit part}. The exact locations of these dark green boxes are summarized in Table \ref{tab: fov sect}. In this format, each lit part has size of $ 150 \times 430 \,\mathrm{pix} $. Hence, any region well outside these boxes, including B1, B1', B2, B2', FFTo, and FFTe, are dark overscan areas. After processing until level 4, it is recommended to crop only the central region of the lit part, i.e., the light green boxes, called the \textit{region of interest} (RoI). This is because the edge pixels are affected by the imperfect flat field and can have arbitrarily large or small pixel values, and can affect the sky estimation in photometric/polarimetric measurements. We normally crop out 10 pixels around the edge (so the final analyzed region is $ 130\times 410 \,\mathrm{pix} $).

\begin{table}[]
\begin{center}
  \begin{tabular}{c|c}
  part  & FITS section \\
  \hline
  J-band, o-ray & [534:683, 306:735] \\ 
  J-band, e-ray & [719:868, 306:735] \\
  H-band, o-ray & [564:713, 331:760] \\
  H-band, e-ray & [741:890, 331:760] \\
  K-band, o-ray & [560:709, 341:770] \\
  K-band, e-ray & [729:878, 341:770]
  \end{tabular}
\end{center}
\caption{The FITS sections of each lit part. The FITS section is in the format of $ [x_\mathrm{min}:x_\mathrm{max}, y_\mathrm{min}:y_\mathrm{max}] $ in 1-indexing (first and last pixels inclusive).}
\label{tab: fov sect}
\end{table}

In this section, the detailed process to obtain the reduced image (levels 1-4) is described. B1, B1', B2, and B2' regions are used to make level 1 data, while FFTo and FFTe regions are used for level 2 data. The FFTo and FFTe regions have the same size ($ 150 \times 1024 \,\mathrm{pix} $), and are set such they overlap with the lit part when left half is shifted by $ +512 \,\mathrm{pix} $ to the right.

\vspace{0.5em}
\subsection{Level 1 and Initial Bad-Pixel Mask} \label{s: lv1}
Since NIC data contains random and additive \textit{vertical pattern} \citep{Ishiguro2011ARNHAO}, locations B1, B1', B2, and B2' were used to remove this vertical pattern for the upper-right, upper-left, lower-right, and lower-left quadrants, respectively. For each column within each of these boxes, a 2-sigma clipped median is calculated. This value is subtracted from all the pixels in the same column in the same quadrant. Then level 1 image is obtained.

On UT 2019-11-21, 20 dark frames with an exposure time of 180 s were obtained for all three detectors (J, H, and K bands). All these images were first reduced to level 1. Then after sigma-clipped median combining these dark frames for each detector, any pixel with values larger than $ \mathrm{max}\{5, D_\mathrm{med} + 5 \sigma_D \} \,\mathrm{ADU} $ or smaller than $ \mathrm{min}\{-5, D_\mathrm{med} - 5 \sigma_D \} \,\mathrm{ADU} $ are flagged as bad-pixel for each detector. Here, $ D_\mathrm{med} $ and $ \sigma_D $ are the 3-sigma clipped median and standard deviation of the combined dark frame, respectively. Secondly, the sigma-clipped sample standard deviation was calculated for each pixel position, and any pixel with high variance was flagged. 

On UT 2020-06-03, 640 dome flat frames with 2 s exposure time were obtained (with varying instrumental rotation angle \texttt{INSROT} and half-wave plate rotator angle \texttt{POL-AGL1}). All frames are reduced to level 1. Then similar to the dark frames, the frames were median combined, and the sample standard deviation maps were generated. Each of the o-/e-ray regions was normalized separately before the combination because the average raw flat count is different in the two regions by up to $ \sim 10\% $. Then the pixels with too low value or too high variance are flagged.

\begin{figure} [tb!]
  \begin{center}
    \includegraphics[width=\linewidth]{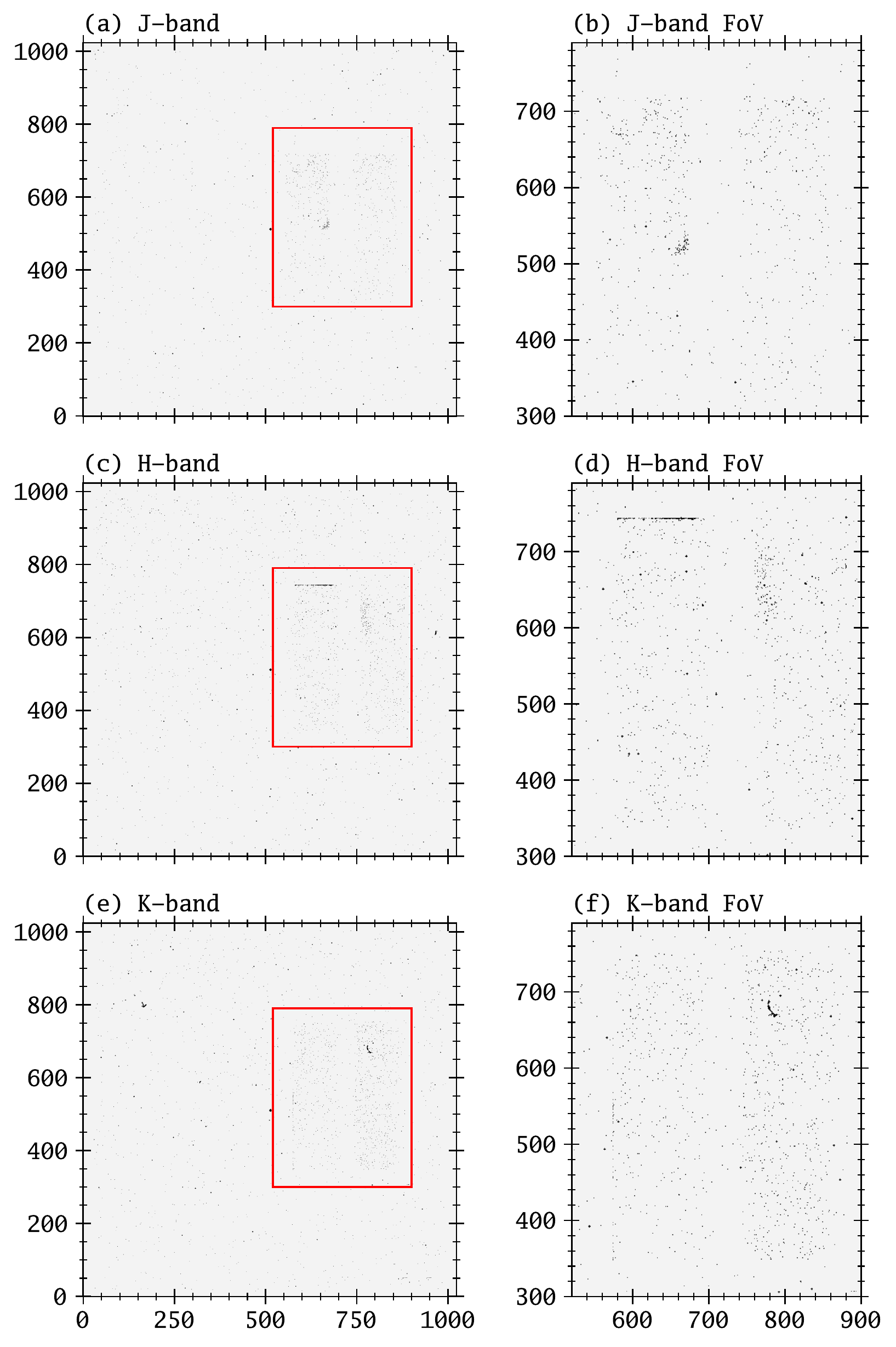}
  \end{center}
  \caption{The distribution of bad pixels in \texttt{imask} frames for J, H, and K bands. The right panels show zoomed-in versions of the red box areas in the left panels, i.e., around the lit parts (green boxes) in Fig. \ref{fig:fignicfitsanatomy}.}
  \label{fig:imask}
\end{figure}

Combining all the flagged pixels from these dark and flat frames, initial bad-pixel masks were generated for each detector (total 3 mask files). These are called the \texttt{imask} in \texttt{NICpolpy}, and shown in Fig. \ref{fig:imask}. As a result, $ \sim 500 $ pixels are masked for each light-green RoI in Fig. \ref{fig:fignicfitsanatomy}. Since the total size after cropping the 10-pixel bezel is $ 130 \times 410 \sim 50,000 \,\mathrm{pix} $, $ \sim 1\% $ of the region is masked. It is found that the position we traditionally put our science target is located where the least number of pixels are masked.

\vspace{0.5em}
\subsection{Nightly Master Mask and Level 2} \label{s: lv2}
Even after removing the vertical pattern, it is known that NIC has an additional quasi-periodic artifact along the diagonal direction (for example, Fig. 28 of \cite{Ishiguro2011ARNHAO} and Fig 5 of \cite{TakahashiJ2018SAG}). Although the specific reason is still unknown, this is likely an additive noise due to an imperfectly shielded external electric signal. After thorough investigations, we could not find a way to remove the noise by tiling columns or rows of the full frame or each quadrant into the 1-dimensional array, and the Fourier transform. Accordingly, this noise is unlikely a perfect periodic wave. Fortunately, it was found that this pattern is quite similar in the left/right quadrants. That is, a direct subtraction of the left part from the right part reduces the pattern. However, this simple subtraction is vulnerable to any cosmic ray hit, amp-glow, newly developed hot-pixels, or even just standard noise (e.g., read-out noise) in the left half of the array. In addition, standard cosmic-ray rejection algorithms (e.g., \cite{2001PASP..113.1420V}) is difficult to be applied directly, because (i) it takes a long time and (ii) the fine-tuning of the parameters for dark (left) area was difficult. Therefore, we applied the following strategy.

As described in Sect. \ref{s: lv1}, we already made the initial pixel mask (\texttt{imask}) from dark frames of UT 2019-11-21. Whenever dark frames are available, those of the longest exposure are used to make a similar pixel mask (called \texttt{dmask}). Then the nightly master mask, \texttt{mmask}, is made by combining those: \texttt{mmask = imask|dmask} (bitwise \texttt{OR} operation). If there are no dark frames at the nith, \texttt{mmask = imask}. This process can remove newly developed bad pixels after UT 2019-11-21\footnote{
  It is also possible to use the dark regions (outside the lit parts) of science exposure frames to find additional bad pixels. This might be useful when there are no nightly dark frames. However, it is less important because the bad pixels in the RoI cannot be retrieved. Hence, it is not implemented yet.
}. We use \texttt{dmask} rather than just subtracting dark, because the absolute value of hot pixels often varies from frame to frame.

Afterwards, similar to, but slightly different from, the previous work \citet{2021A&A...653A..99T}, the left half of the $ 1024 \times 1024 \,\mathrm{pix} $ frame is used to estimate and remove the periodic pattern (also called the \textit{Fourier pattern} in \texttt{NICpolpy}). To minimize the computation time, this pattern inference is made on FFTo/FFTe regions only rather than the full left-half frame. The pixels within FFTo/FFTe regions that are flagged in \texttt{mmask} are interpolated. This interpolation is similar to \texttt{IRAF.PROTO.FIXPIX}, prioritizing the vertical direction over the horizontal one (the priority can be modified by the user), implemented for $ N $-dimensional array by using \texttt{scipy}. 

In addition, the pixels that are likely cosmic-ray, newly developed bad-pixel, or high-noise artifacts are replaced by the median-filtered value. Three parameters are calculated from the image in this algorithm, called the Median-filter Bad-Pixel Masking (MBPM):
\begin{equation}\label{ea: mbpm}
    C_1 = \mathcal{I} - \mathcal{M} ~,\\
    C_2 = \mathcal{I} / \mathcal{M} ~,\\
    C_3 = (\mathcal{I} - \mathcal{M})/\sigma(\mathcal{I}) ~.
\end{equation}
Here, $ \mathcal{I} $ is the original image, $ \mathcal{M} $ is the median filtered image, and $ \sigma(\mathcal{I}) $ is the sigma-clipped standard deviation at certain user-tunable region(s) of $ \mathcal{I} $. In this work, the pixels with $ |C_1| > 5 \, [\mathrm{ADU}]$ or $ C_2 \notin [0.5, 2] $ or $ |C_3| > 5 $ are flagged and replaced with the median-filtered value. The regions for $ \sigma(\mathcal{I}) $ is [50:450, 100:900] by default in the FITS section format. It is not iterated multiple times. The median-filter size was $ 5 \times 5 \,\mathrm{pix} $, and a very small real number is added to $ \mathcal{M} $ to avoid zero-division. This MBPM algorithm is orders of magnitude quicker than traditional cosmic-ray rejection algorithms (e.g., \cite{2001PASP..113.1420V}) while removing artifacts/hot-pixels/cosmic-rays reasonably well and is easily applicable to the dark region without fine-tuning of the parameters \footnote{
  MBPM is not intended to detect cosmic rays in the presence of celestial objects. Unlike algorithms that use edge detection and convolutions to find cosmic rays, MBPM only uses the deviation from the median filtered image to find outliers without any additional convolution or edge detection. Thus, MBPM may easily flag stellar objects. MBPM is intended to quickly find bad pixels or cosmic rays in dark frames or sky regions where no point source is expected. MBPM is separately available as a function at \texttt{NICpolpy.ysfitsutilpy4nicpolpy.medfilt\_bpm}.
}.

After smoothing out these artifacts, a fast Fourier transform (FFT) is applied to each column in the FFTo and FFTe regions. Then the components with wavelength $ > 100 \,\mathrm{pix} $ are extracted, so that two pattern maps with sizes $ 150 \times 1024 \,\mathrm{pix} $ are made. These pattern maps are subtracted from the right part, overlapping with o-ray and e-ray RoI, respectively. This image undergoes an additional vertical pattern subtraction, as in level 1, to remove any remaining vertical pattern (likely smaller than $ 0.5 \,\mathrm{ADU}$).

The resulting image is denoted as \textit{level 2}. So far, it is found that this process reduces the periodic pattern, especially for the frames having larger contamination ($ \lesssim \pm 5 $ ADU, see, Fig. \ref{fig:fignicfitsanatomy}c). We confirmed that this process does not change the final Stokes' parameter values while greatly reducing background fluctuation. These reduction processes would be beneficial for extended source studies, e.g., the Moon, earth shine \citep{2021A&A...653A..99T} or comet polarimetry.

\vspace{0.5em}
\subsection{Master Flat and Level 3 and 4}
We found no clear and severe difference in the combined flats for different \texttt{POL-AGL1}. We hence did not split the master flat based on the rotator angles (Sect. \ref{s: flat} and supporting material in Appendix \ref{s: sm}). The master flat was generated by the following steps: 
\begin{enumerate}
    \item sum combines each polarimetric set (four frames at four \texttt{POL-AGL1}),
    \item normalize each of the sum images based on the sigma-clipped median in the o-ray RoI,
    \item median combines these by sigma-clipping.
\end{enumerate}
Because the flat counts are not normalized separately for o-/e-ray regions, the final flat has slightly different values for the o-/e-ray regions ($\sim 1$ and $\sim 1.1$, respectively). This difference is not important because NIC samples the flux twice per each Stokes' parameter (e.g., \texttt{POL-AGL1} of 0 and $45^\circ$ for $q$) to cancel out this offset effect.

After removing the instrumental noises (vertical and Fourier patterns), a standard preprocessing, including dark subtraction and flat fielding, was done. Using \texttt{mmask}, some pixels are interpolated. Then each frame is split into two frames, o-ray and e-ray (each $ 150 \times 430 \,\mathrm{pix} $). This is \textit{level 3} data. If the user demands, blank sky frames are extracted and combined after scaling to make a master fringe frame. Otherwise, blank sky frames will be discarded at this stage.

Finally, fringe subtraction and cosmic-ray rejection are done on each image, saving the final result as \textit{level 4}. Fringe is approximated as an additive feature in imaging observation. In \texttt{NICpolpy}, fringe correction is implemented either by scaling the normalized sky fringe or direct subtraction. The latter is sometimes erroneous because the sky brightness (and even the fringe pattern) can change rapidly. As it will be justified later in future work, the fringe subtraction is implemented but omitted in real data reduction. Each fringe frame is divided by the sigma-clipped average pixel value to eliminate the overall sky brightness change. After this normalization, frames are combined to make a master fringe frame (for the fine tuning, see Sect. \ref{sss: cell9}). Since the fringe pattern is well known to be variable over time, \texttt{NICpolpy} automatically combines fringe based on time stamps, too (Sect. \ref{sss: cell8}). The HWP angle, \texttt{POL-AGL1}, is ignored when combining master fringe because we found that separating fringe by HWP angle does not improve the final results. For the cosmic-ray rejection, \texttt{astroscrappy} \citep{astroscrappy_1_0_5}, which implements \texttt{L.A.Cosmic} \citep{2001PASP..113.1420V}, is used (version 1.1.1). After the investigation, the parameters similar to \texttt{L.A.Cosmic} are found to be able to remove cosmic rays efficiently.

The blank sky (fringe) frame can approximately be regarded as a dark frame, especially for short exposures and when the sky is dark. This, however, is vulnerable to the change in sky level, spatial polarization patterns in the sky, and typical pixel noise (readout noise and Poisson noise) in sky frames because we have few sky frames. Therefore, this function has not yet been implemented to \texttt{NICpolpy}. Another approach is to conduct a dithering observation to construct a fringe map. This technique is also difficult for NIC because of the small field-of-view, especially when sky level and/or fringe pattern vary quickly relative to the exposure time. Finally, it can be shown that the fringe pattern has a negligible effect on Stokes' parameter determination for beam-splitter type aperture polarimetry when the effect of fringe is a few-percent level of the aperture sum and the true polarization degree of the target is $ \lesssim \mathrm{few}\% $ (Bach et al. in prep.). Thus, we ignore fringe in this work.

\section{Code Usage and Implementation}\label{s: code usage}
This section describes the practical usage of the package. Since \texttt{NICpolpy} is registered to the Python Package Index (PyPI), the installation is invoked once typing
\begin{equation}
    \texttt{\$ pip install NICpolpy} 
\end{equation}
on the terminal\footnote{
    The terminal must have \texttt{pip} installed: \texttt{https://pypi.org/project/pip/}
}.
The development version is available via GitHub\footnote{
    \texttt{https://github.com/ysBach/NICpolpy}
}. After the installation, calibration frames \texttt{imask} and flats are required. They are available in a separate GitHub repository (Appendix \ref{s: sm}). 

First, the simplest way to use \texttt{NICpolpy} is briefly discussed. Then the output FITS files and log files are explained. Finally, the core algorithmic implementations of reduction processes (Sect. \ref{s: dataprep}) are described step by step, as well as intermediate output files. Although the detailed usage can include many more arguments (and may change in future developments), here we describe the core ideas and basic usage that will suffice most of the users' needs.

\vspace{0.5em}
\subsection{Basic Usage}
The most recent development of \texttt{NICpolpy} as of writing is happening on Python 3.10 with macOS 12. However, since each part of the code was written OS-agnostic, it will work without any problem on any recent major OS (Windows, macOS, Linux) and python versions (3.8 or later). 

First, import packages and initialize the reducer by giving the names of the directories where the original data and calibration data are stored\footnote{
    This example is also provided in Appendix \ref{s: sm}.
}:
\begin{footnotesize}
\begin{verbatim}
# import nicpolpy package
import nicpolpy as nic

# Do not print useless warnings
import warnings
from astropy.utils.exceptions import AstropyWarning
warnings.filterwarnings('ignore', 
    append=True, category=AstropyWarning)
    
# Cell 0: Initialize the reducer
npr = nic.NICPolReduc(
    name="SP_20190417",
    inputs="_original_32bit/190417/raw/*.fits",
    mflats="cal-flat_20180507-lv1/*.fits",
    imasks="masks/*.fits",
    verbose=1
)
\end{verbatim}
\end{footnotesize}
For the reducer, the variable \texttt{name} is the prefix that will be used for result files (see Sect. \ref{ss: fits name} and \ref{ss: log}). The argument \texttt{inputs}, \texttt{mflats} and \texttt{imasks} indicates the regex-style for input FITS files, master flat files, and \texttt{imask} files, relative to the current working directory, respectively. The next steps are
\begin{footnotesize}
\begin{verbatim}
  # Cell 1: Planner for master dark combination
  #   5 sec
  _ = npr.plan_mdark()
  # Cell 2: Combine dark & planner for master mask
  #   11 sec
  _ = npr.comb_mdark_dmask()
  _ = npr.plan_mmask()
  # Cell 3: Make master mask
  #   << 1 sec
  _ = npr.comb_mmask()
  # Cell 4: Make lv1 & planner for lv2
  #   35 sec/456 FITS
  _ = npr.proc_lv1() 
  _ = npr.plan_lv2()
  # Cell 5: Make lv2 & planner for lv3 
  #   4 min / 357 FITS
  _ = npr.proc_lv2() 
  _ = npr.plan_lv3()
  # Cell 6: Make lv3
  #   2 min
  _ = npr.proc_lv3()
  # Cell 7: Planner for lv4 of fringe frames
  #   1 min
  _ = npr.plan_ifrin()
  # Cell 8: Planner for master fringe
  #   35 sec
  _ = npr.plan_mfrin()
  # Cell 9: Make master fringe
  #   few sec
  _ = npr.comb_mfrin()
  # Cell 10: Planner for lv4
  #   few sec
  _ = npr.plan_lv4(add_mfrin=False)
  # Cell 11: Make lv4
  #   2 min
  _ = npr.proc_lv4()
\end{verbatim}
\end{footnotesize}
The code lines indicated by different \texttt{Cell} numbers are recommended to run after checking the results from the previous \texttt{Cell}. 
Note that, except for the input file directories, the user does not have to specify any specific parameters and quickly obtain the final level 4 files. For specific arguments for fine-tuning, see Sect. \ref{ss: imp details}. The comments with \texttt{xx min} or \texttt{xx sec} are the time spent on Mac Book Pro with m1Pro chip (8 performance, 2 efficiency cores with 16 GB RAM).

From now on, the following simplifying notations are used: 
\begin{eqnarray*}\label{eq: summ plan}
  \textrm{log directory} = \texttt{\_\_logs/<name>}~, \\
  \textrm{level~} \texttt{<i>} \mathrm{~directory} = \texttt{\_lv<i>/<name>}~,\\
  \texttt{summ(X)} = \texttt{<name>\_summ\_X.csv}~, \\
  \texttt{plan(X)} = \texttt{<name>\_plan-X.csv}~, 
\end{eqnarray*}
where \texttt{name} is the argument given in \texttt{Cell 0}, \texttt{<i>} is an integer from 1 to 4, and \texttt{X} is any name. For example, \texttt{plan(MDARK)} in this specific example means the file \texttt{SP\_20190417\_plan-MDARK.csv}, inside the corresponding \textit{log directory}, \texttt{\_\_logs/SP\_20190417/}. After running the code \texttt{cell}'s above, multiple outputs are generated. Most importantly, the FITS files are saved in the corresponding level directory, \texttt{\_lv<i>/<name>}. Also, the log information and all the calibration files are saved in the log directory, \texttt{\_\_logs/<name>}. The contents are described in the following subsections.

\vspace{0.5em}
\subsection{Output: FITS Files and Level Directories} \label{ss: fits name}
Any FITS file written by \texttt{NICpolpy} has the following naming convention:
\begin{footnotesize}
\begin{verbatim}
  <FILTER>_<System YYYYMMDD>
  _<COUNTER:04d>_<OBJECT>_<EXPTIME:.1f>
  _<POL-AGL1:04.1f>_<INSROT:+04.0f>
  -PROC-{PROCESS}_{QU}{SETID:03d}[_{OERAY}].fits
\end{verbatim}
\end{footnotesize}
Each part means
\begin{itemize}
  \item \texttt{FILTER}: Lower-cased filter (\texttt{j}, \texttt{h}, or \texttt{k}).
  \item \texttt{YYYYMMDD}: The date when this image was obtained. Same as the original data saved by NIC.
  \item \texttt{COUNTER}: The image counter in 4 digits with leading 0's. It starts from 1.
  \item \texttt{OBJECT}: The object name as in FITS header key \texttt{OBJECT}.
  \item \texttt{EXPTIME}: The exposure time in seconds as in FITS header key \texttt{EXPTIME}.
  \item \texttt{POL-AGL1}: The HWP rotator angle as in FITS header key \texttt{POL-AGL1}. One of \texttt{00.0}, \texttt{45.0}, \texttt{22.5} or \texttt{67.5}. It is \texttt{xxxx} if there is no \texttt{POL-AGL1} in the header.
  \item \texttt{INSROT}: The instrument rotator angle at exposure as in FITS header key \texttt{INSROT}. 
  \item \texttt{PROCESS}: The history of processing. It can contain \texttt{D} (dark subtraction), \texttt{F} (flat-fielding), \texttt{T} (trimmed), \texttt{C} (cosmic-ray rejection), \texttt{R} (fringe subtraction), \texttt{P} (bad pixel interpolation), \texttt{v} (vertical pattern removal), and \texttt{f} (Fourier pattern removal). The processing order is from left to right.
  \item \texttt{QU}: Either \texttt{q} or \texttt{u}, which means whether the frame will be used for the raw $ q $ or $ u $ parameter calculation. Former is when \texttt{POL-AGL1} is either \texttt{00.0} or \texttt{45.0}, and the latter is when either \texttt{22.5} or \texttt{67.5}.
  \item \texttt{SETID}: The 3-digit integer counter with leading 0's for the group (set) id for polarimetric measurement. One set consists of four exposures with different \texttt{POL-AGL1} angles, and the counter increases after the cycle. It starts from 1.
  \item \texttt{OERAY}: Either \texttt{o} or \texttt{e}, indicating whether this is o-ray or e-ray RoI (only for level 3 and 4).
\end{itemize}

The resulting files will be saved in the corresponding level directories. At each level directory, a sub-directory for thumbnails of each FITS frame called \texttt{thumbs/}, is generated when making the planner for the level \texttt{<i+1>} (see Sect. \ref{ss: imp details}). There can be other sub-directories, \texttt{\_lv1/<name>/dark}, if there are nightly darks, and \texttt{\_lv4/<name>/frin}, if there are fringe (blank sky) frames.

\vspace{0.5em}
\subsection{Output: Log Directory (\texttt{\_\_logs/})} \label{ss: log}
\begin{figure*} [tb!]
  \begin{center}
    \includegraphics[width=0.7\linewidth]{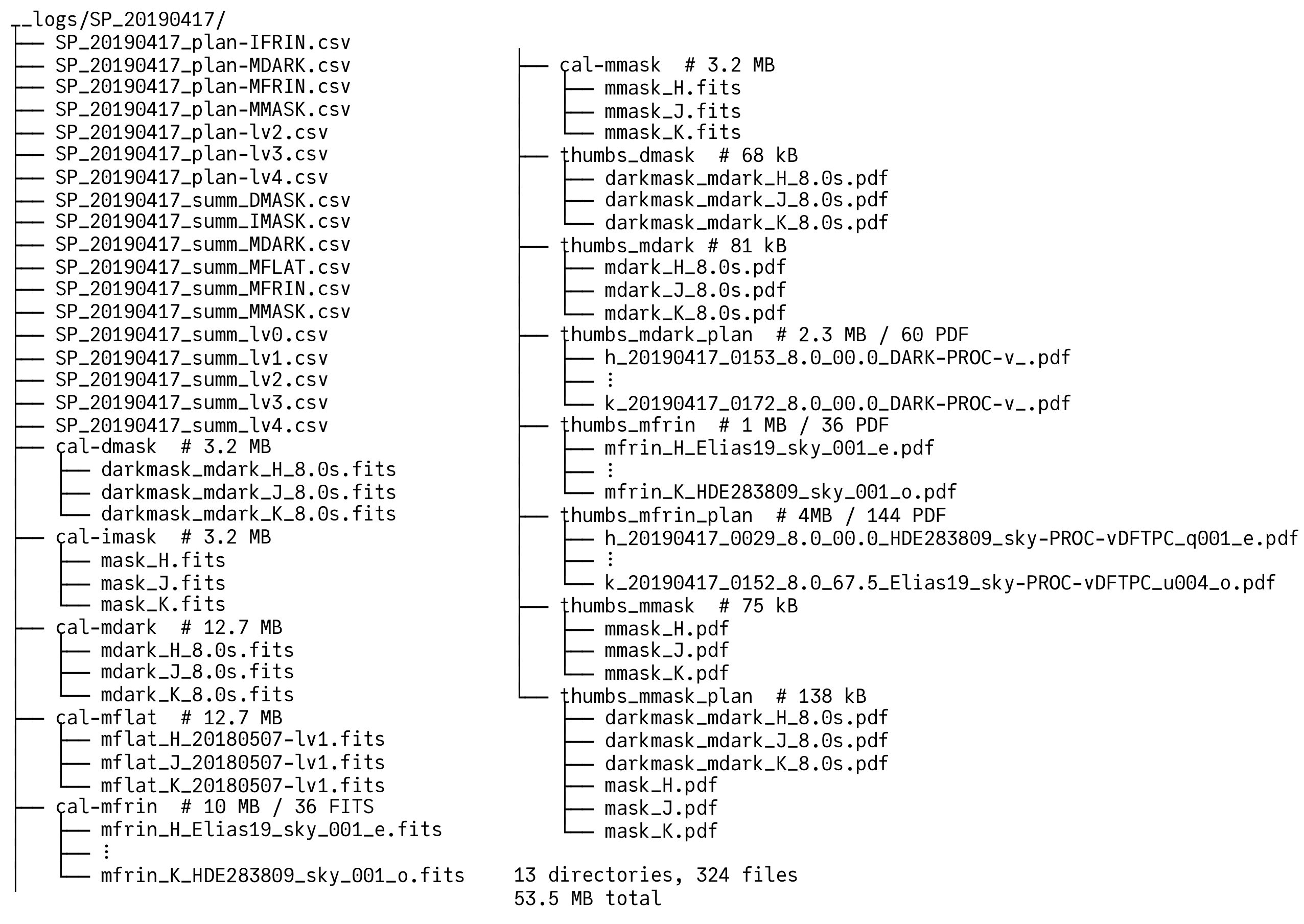}
  \end{center}
  \caption{The final tree structure of the log directory, \texttt{\_\_logs/SP\_20190417}, as an example. Total 13 directories, 324 files, and 53.5 MB. The size of each sub-directory is indicated as comments.}
  \label{fig:tree}
\end{figure*}

After initializing the reducer (\texttt{Cell 0}), \texttt{\_\_logs/<name>} directory is made. This is called the log directory, and its tree structure is shown in Fig. \ref{fig:tree}. This is made to archive virtually all necessary information to guarantee the reproducibility of the user's work. The calibration files (master flat and mask files) are copied and saved, and all the thumbnail images related to calibration will be saved here. 
Inside here are multiple CSV files and directories:
\begin{itemize}
  \item \texttt{cal-xxxxx/}: Contains calibration FITS frames.
  \item \texttt{thumbs\_xxxx/}: The thumbnails with certain statistics of each frame for quick-look.
  \item \texttt{summ(X)}: The summary of FITS frames for each level or master calibration frames, e.g., \texttt{SP\_20190417\_summ\_lv0.csv}.
  \item \texttt{plan(X)}: The planner before proceeding to each level or master calibration frame generation, e.g., \texttt{SP\_20190417\_plan-MDARK.csv}.
\end{itemize}

\vspace{0.5em}
\subsection{Implementation Details} \label{ss: imp details}
Here, the details of implementation and the outputs from each \texttt{Cell} number are described. As most of the tools do, \texttt{NICpolpy} also may undergo quick and breaking changes when necessary. So for the most details, such as tuning output thumbnail files, which do not affect the scientific outcomes, it is recommended to look at the source code. Also, there are multiple arguments or hard-coded parts that are useful merely for debugging purposes, which are not described here.

\vspace{1em}
\subsubsection{Cell 0: Initialization}\label{sss: cell0}
After the initialization, the log directory and level 1-4 directories are automatically generated. Then inside the log directory, \texttt{cal-imask/} and \texttt{cal-mflat/}, based on \texttt{imasks} and \texttt{mflats} arguments, are made. These are not made if arguments are not given (\texttt{None}). Also, the summary of level 0 files are made as \texttt{summ(lv0)} inside the log directory. This contains some selected header information from each FITS file. If \texttt{imasks} or \texttt{mflats} are given, corresponding summary files are also made: \texttt{summ(MFLAT)} and \texttt{summ(IMASK)}. These two contains little information, and made only for the consistency.

\vspace{1em}
\subsubsection{Cell 1: Master Dark Planner}\label{sss: cell1}
The important default arguments are:
\begin{footnotesize}
\begin{verbatim}
npr.plan_mdark(
    method="median",
    sigclip_kw=dict(sigma=2, maxiters=5),
    fitting_sections=None,
)
\end{verbatim}
\end{footnotesize}

\begin{figure*} [tb!]
  \begin{center}
    \includegraphics[width=0.7\linewidth]{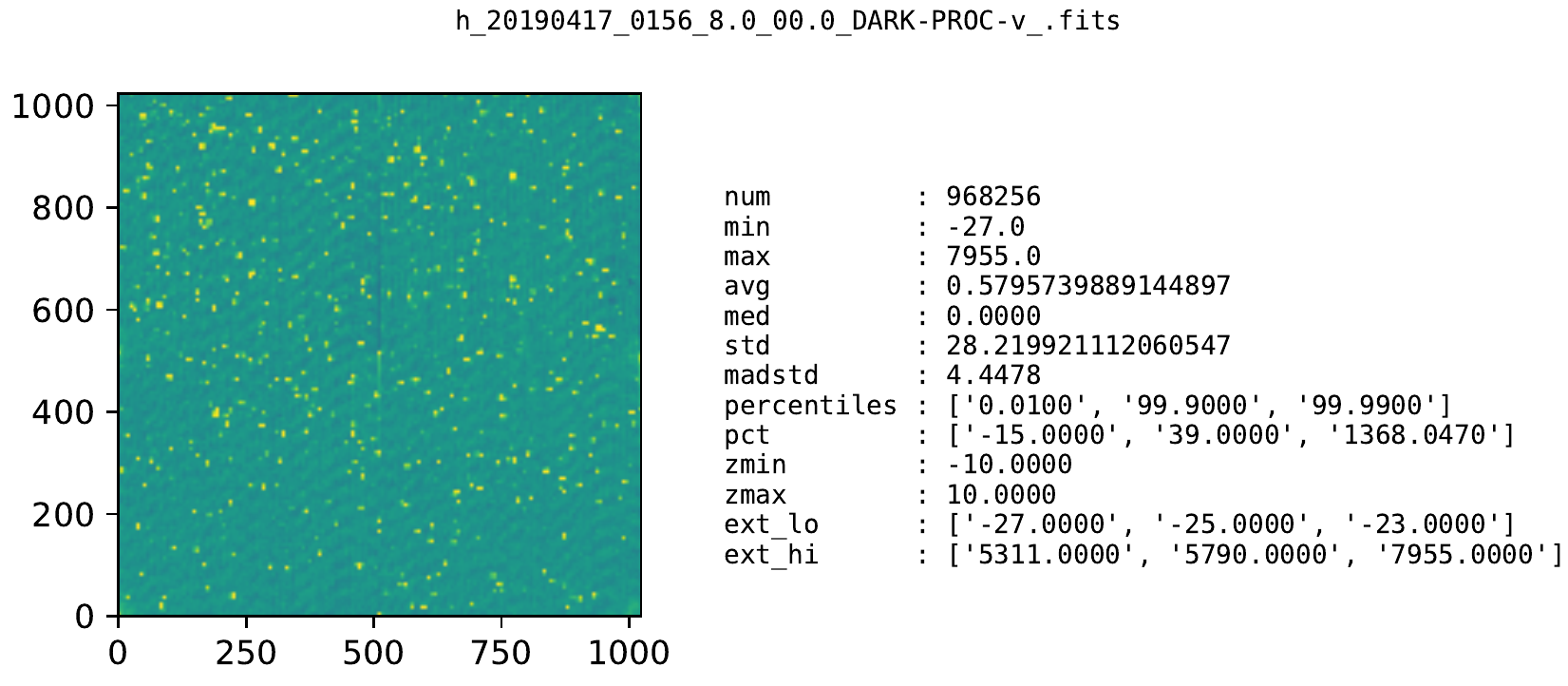}
  \end{center}
  \caption{An example of saved thumbnail images in \texttt{\_\_logs/<name>/thumbs\_mdark\_plan}. It is a DARK frame with 8 s exposure in H-band, as shown in the title. The statistics are based on pixels ignoring 20-pixel edge bezels. The statistics are the number of pixels, minimum, maximum, average, median, sample standard deviation, and median-absolute-deviation (MAD) derived standard deviation of pixel values. \texttt{pct} are the corresponding percentile pixel values specified by \texttt{percentiles}. \texttt{zmin} and \texttt{zmax} are the famous zscale minimum/maximum values. \texttt{ext\_lo} and \texttt{ext\_hi} are the top three pixel values in the lowest and highest regimes, respectively.}
  \label{fig:mdarkplan}
\end{figure*}

Among level 0 files, it finds those with \texttt{OBJECTS = "DARK"} in the header. If there is any nightly dark frames, they undergo vertical pattern removal (pattern is estimated in the regions \texttt{fitting\_sections}, and using the statistic \texttt{method}, obtained by sigma-clipping with \texttt{sigclip\_kw}), and saved into \texttt{dark/} inside level 1 directory. The planner \texttt{plan(MDARK)} is made. This planner contains file name, timestamp, filter, and exposure time information. There is a column named \texttt{REMOVEIT}, and if the user fills here with any value other than 0, the frame will be ignored in the master dark combination. For each image, a thumbnail is made with some statistical information and saved to \texttt{thumbs\_mdark\_plan/} in the log directory. An example is shown in Fig. \ref{fig:mdarkplan}. The user may look at the thumbnails and pinpoint strange dark frames to fill in \texttt{REMOVEIT}. For example, at one night, we could find multiple dark frames have abnormally high median values of $ > 5 \,\mathrm{ADU} $ only at K-band based on these thumbnails. Such frames can be removed before making the master dark in the next \texttt{Cell}.

\vspace{1em}
\subsubsection{Cell 2: Master Dark and Master Mask Planner}\label{sss: cell2}
The important default arguments are:
\begin{footnotesize}
\begin{verbatim}
npr.comb_mdark_dmask(
    combine_kw=dict(combine="med", reject="sc", sigma=3),
    dark_thresh=(-10, 50),
    dark_percentile=(0.01, 99.99),
)
npr.plan_mmask()
\end{verbatim}
\end{footnotesize}

\begin{figure*} [tb!]
  \begin{center}
    \includegraphics[width=0.7\linewidth]{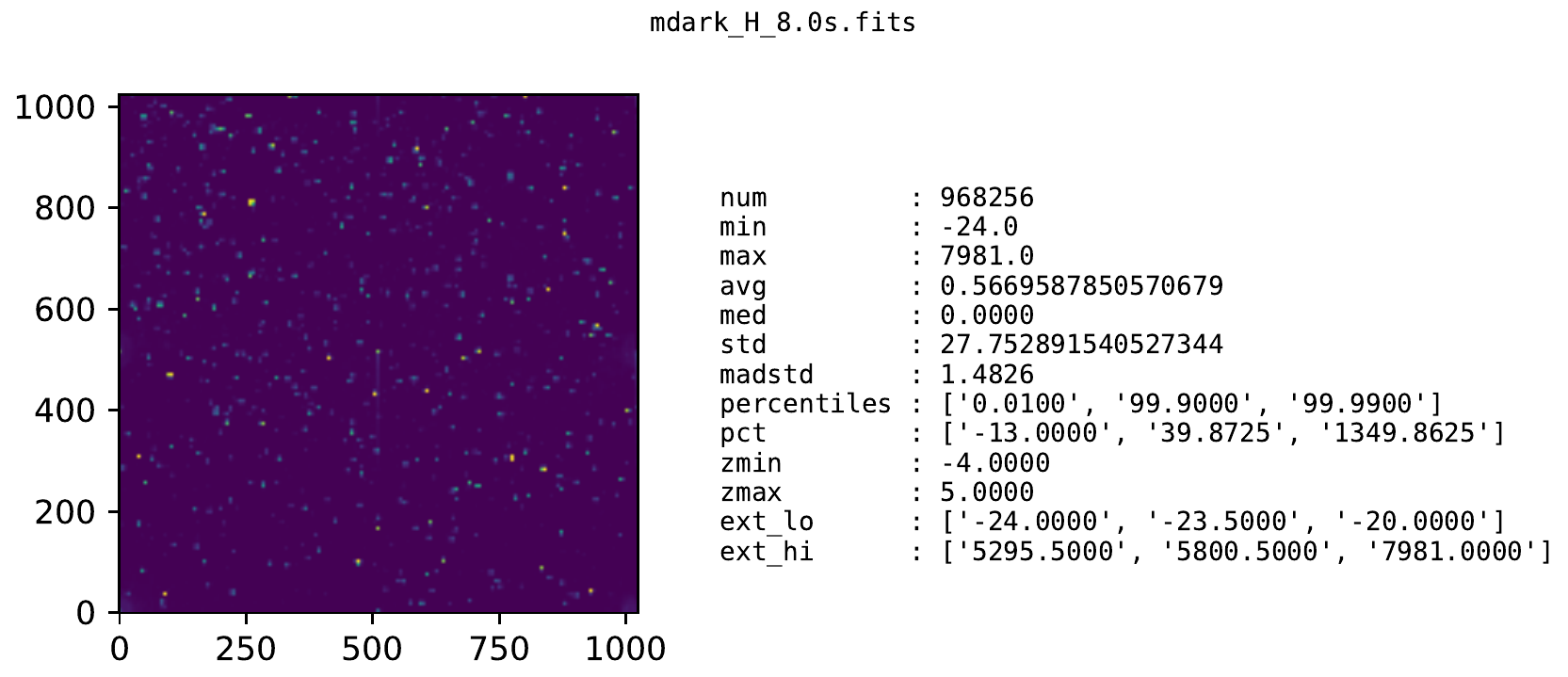}
  \end{center}
  \caption{An example of saved thumbnail images in \texttt{\_\_logs/<name>/thumbs\_mdark/}. It is the master dark frame with 8 s exposure in H-band, as shown in the title. The statistics are as in Fig. \ref{fig:mdarkplan}.}
  \label{fig:mdark}
\end{figure*}

The master dark is then made by median combined with 3-sigma clipping (tunable by \texttt{combine\_kw}). The resulting master darks are saved in \texttt{cal-mdark/}, while the thumbnails are saved in \texttt{thumbs\_mdark/} in the log directory. An example of thumbnails is shown in Fig. \ref{fig:mdark}. The summary of master darks is also saved as \texttt{summ(MDARK)}. For each master dark frame, any pixel that are outside $ [-10, 50] \,\mathrm{ADU} $ range, or $ [0.01, 99.99] $ percentile range are flagged (tunable by \texttt{dark\_thresh} and \texttt{dark\_percentile}, respectively). 

The generated intermediate \texttt{dmask} files and their thumbnails are saved in \texttt{cal-dmask/} and \texttt{thumbs\_dmask/}, and summary file as \texttt{summ(DMASK)} in the log directory, respectively. Each \texttt{dmask} frame is in the format of \texttt{uint8} (unsigned 8-bit integer). Pixel values are calculated by summing the following bits:
\begin{itemize}
  \item 00000010 = 2: dark above upper threshold,
  \item 00000100 = 4: dark above upper percentile,
  \item 00001000 = 8: dark below lower threshold,
  \item 00010000 = 16: dark below lower percentile,
\end{itemize}
so that the minimum and maximum pixel values are 2 and 30, respectively. 

Then the planner for nightly \texttt{mmask} is made as \texttt{plan(MMASK)} file, while the thumbnails for \texttt{imask} and \texttt{dmask} are saved in \texttt{thumbs\_mmask\_plan/} inside the log directory, respectively. The user can remove certain \texttt{imask} or \texttt{dmask} for making \texttt{mmask}, by changing \texttt{REMOVEIT} column in \texttt{plan(MMASK)} as before.

\vspace{1em}
\subsubsection{Cell 3: Master Mask}\label{sss: cell3}
The important default arguments are:
\begin{footnotesize}
\begin{verbatim}
npr.comb_mmask(combine_kw=dict(combine="sum", reject=None))
\end{verbatim}
\end{footnotesize}

It is a simple step: Just summing the original \texttt{imask} and \texttt{dmask} to make \texttt{mmask}. The user should check if \texttt{mmask} is reasonable at this step. If not, it is likely that a few dark frames had an abnormality. Any pixels with non-zero value in \texttt{mmask} will be regarded as bad-pixels. Final \texttt{mmask}, their thumbnails, and summary are saved in \texttt{cal-mmask/}, \texttt{thumbs\_mmask/}, and \texttt{summ(MMASK)} inside the log directory, respectively. It is recommended not to change any argument here.

\vspace{1em}
\subsubsection{Cell 4: Make lv1 and lv2 Planner}\label{sss: cell4}
The important default arguments are:
\begin{footnotesize}
\begin{verbatim}
npr.proc_lv1(
    method="median",
    sigclip_kw=dict(sigma=2, maxiters=5),
    fitting_sections=None,
)
npr.plan_lv2(maxnsat_oe=5, maxnsat_qu=10)
\end{verbatim}
\end{footnotesize}

Running \texttt{proc\_lv1} in this cell will process all frames (except for dark) to remove the vertical pattern (arguments same as the master dark planner: the pattern is estimated in the regions \texttt{fitting\_sections}, and using the statistic \texttt{method}, obtained by sigma-clipping with \texttt{sigclip\_kw}), and save them to the level 1 directory with the filename described in Sect. \ref{ss: fits name}. Then a summary file \texttt{summ(lv1)} will be saved in the log directory. Since NIC loses linearity at around 8000 ADU \citep{Ishiguro2011ARNHAO}, it flags if a frame has many saturated pixels when making \texttt{summ(lv1)}. For each level 1 FITS file, the number of saturated pixels in the o-/e-ray lit parts are counted. These numbers are written in \texttt{NSATPIXO} and \texttt{NSATPIXE} columns, respectively, in both the \texttt{summ(lv1)} and \texttt{plan(lv2)} files. 

When making \texttt{plan(lv2)}, the \texttt{REMOVEIT} code value is calculated by the following logic:
\begin{itemize}
  \item [\texttt{0}]: Saturated pixels have minor effects.
  \item [\texttt{1}]: Number of saturated pixels is larger than \texttt{maxnsat\_oe} (default 5) at one or more of o-ray or e-ray lit parts in this FITS file.
  \item [\texttt{2}]: The total number of saturated pixels at the four lit parts in $ \texttt{POL-AGL1} = \theta \,\mathrm{and}\, \theta + 45^\circ $ frames is larger than \texttt{\texttt{maxnsat\_qu}} (default 10). Both FITS files will have \texttt{REMOVEIT} added by 2.
\end{itemize}
Those with code \texttt{1} will result in an unreliable aperture sum value of at least one of o-/e-ray. Those with \texttt{2} will result in unreliable Stokes' parameters. The final code is the sum of these codes. Note that any frame with \texttt{REMOVEIT!=0} will be ignored from the next (level 2) procedure. That is, \texttt{NICpolpy} automatically ignores frames with many saturated pixels. 

\begin{figure*} [tb!]
  \begin{center}
    \includegraphics[width=0.7\linewidth]{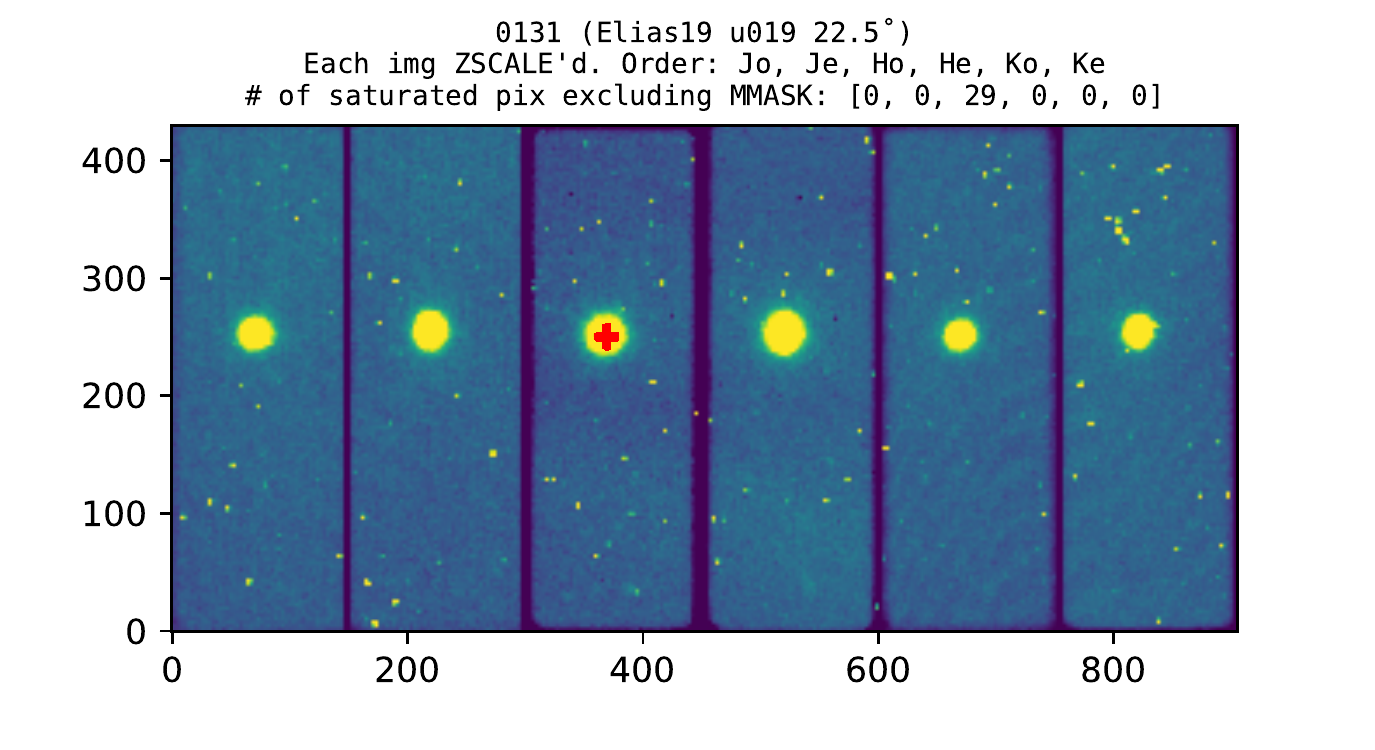}
  \end{center}
  \caption{An example of saved thumbnail images in \texttt{\_lv1/<name>/thumbs/}. The J, H, and K band o-/e-ray images are stacked into one image. In the title, the first line is ``\texttt{<COUNTER> (<OBJECT> <QU><SETID> <POL-AGL1>$ ^\circ $)}'' (see Sect. \ref{ss: fits name} for description). The last line shows the number of saturated pixels, except for those that have already masked in \texttt{mmask}. The red crosses in the image indicate the saturated location. Many pixels in H-band o-ray from the target are saturated, and the resulting Stokes' parameter will be unreliable in the H-band.}
  \label{fig:lv1}
\end{figure*}

After running it, thumbnail files will also be saved in \texttt{thumbs/} inside the level 1 directory. An example is shown in Fig. \ref{fig:lv1}. Using the thumbnail, the user can modify the auto-generated \texttt{REMOVEIT} column in \texttt{plan(lv2)} to select which frames to discard or use.

\vspace{1em}
\subsubsection{Cell 5: Make lv2 and lv3 Planner}\label{sss: cell5}
The important default arguments are:
\begin{footnotesize}
\begin{verbatim}
npr.proc_lv2(
    cut_wavelength=100,
    bpm_kw=dict(
        size=5,
        med_sub_clip=[-5, 5],   # C_1 in Eq (1)
        med_rat_clip=[0.5, 2],  # C_2 in Eq (1)
        std_rat_clip=[-5, 5],   # C_3 in Eq (1)
        logical="and",
        std_model="std",
        std_section="[50:450, 100:900]",
        sigclip_kw=dict(sigma=3., maxiters=50, std_ddof=1)
    ),
)
npr.plan_lv3(use_lv1=False)
\end{verbatim}
\end{footnotesize}
Using level 1 data, level 2 data is processed. The user has large degrees of freedom, but the most important argument is the minimum wavelength of the Fourier pattern (\texttt{cut\_wavelength}) in pixel unit. By default, it is set as 100 pixels. The MBPM algorithm parameters for finding bad pixels can be tuned, too. However, parameters for MBPM do not result in a large difference unless they are tuned such that hot pixels or cosmic-rays are missed. \texttt{proc\_lv2} will generate \texttt{summ(lv2)} in the log directory.

\begin{figure*} [tb!]
  \begin{center}
    \includegraphics[width=0.7\linewidth]{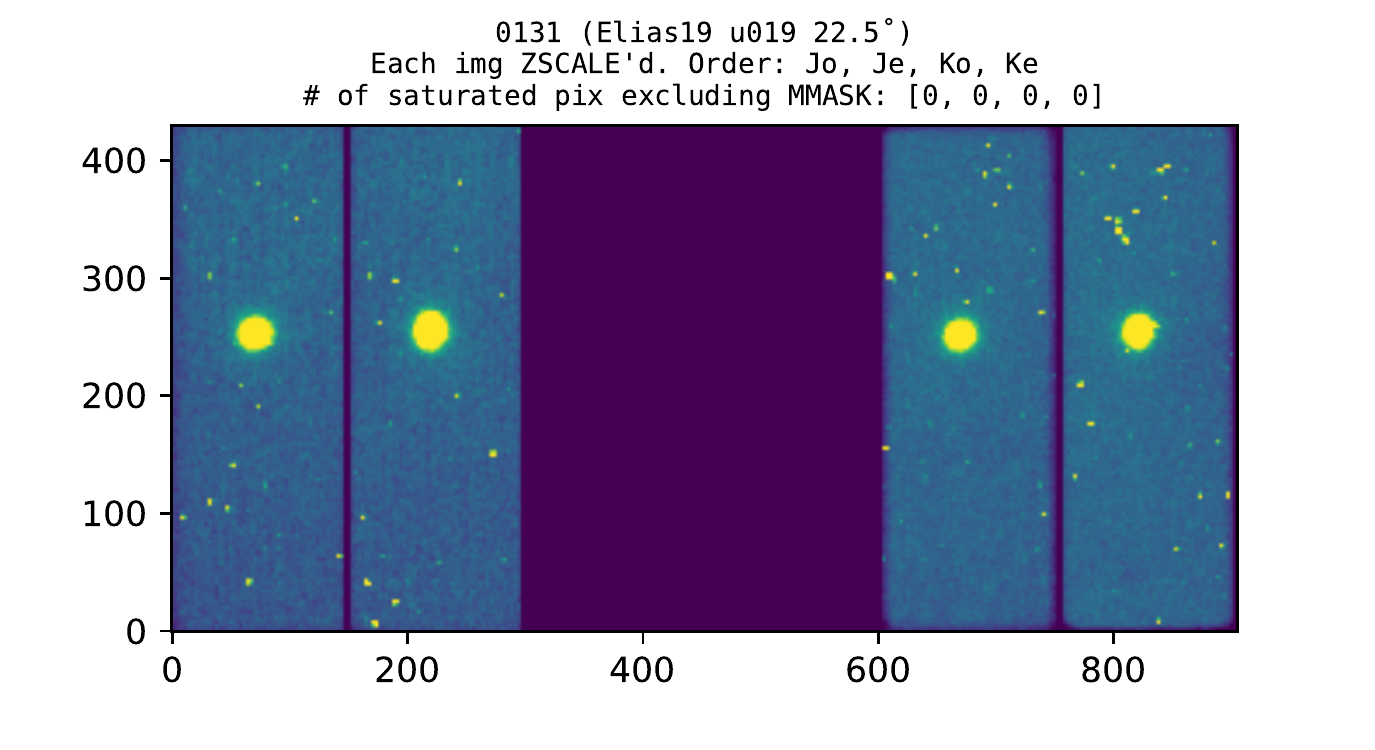}
  \end{center}
  \caption{An example of saved thumbnail images in \texttt{\_lv2/<name>/thumbs/}, identical to Fig. \ref{fig:lv1}. H-band images are removed (non-zero \texttt{REMOVEIT} in \texttt{plan(lv2)}). Note that the Fourier pattern is reduced compared to Fig. \ref{fig:lv1}.}
  \label{fig:lv2}
\end{figure*}

The planner for level 3 will be made after \texttt{plan\_lv3}. Based on the exposure time and filter information in the header, corresponding master dark and flat frames will be found from the log directory, and will be added as \texttt{DARKFRM} and \texttt{FLATFRM} columns in \texttt{plan(lv3)} file. The thumbnails of level 2 frames will be saved in \texttt{thumbs/} in the level 2 directory (Fig. \ref{fig:lv2}). 

The generation of level 2 can take quite a long time. The bad pixel finding and FFT are rather quick, while most of the time is spent for interpolating the masked pixels by \texttt{mmask}. It is possible that this Fourier pattern removal does not work for some cases we could not expect, or the cause for the Fourier pattern can be removed in the future. Also, because the final Stokes' parameter values are not changed much from this step\footnote{
  It is qualitatively understandable because NIC has a large FWHM in pixel unit. Because of that, the aperture size should be set large ($ \sim 40\,\mathrm{pix} $), and the positive and negative Fourier pattern offsets within the aperture will roughly cancel out.
},
the user may completely ignore level 2 (remove \texttt{proc\_lv2} line) and just proceed making level 3 from level 1 data, by \texttt{use\_lv1=True}.

\vspace{1em}
\subsubsection{Cell 6: Make lv3}\label{sss: cell6}
The important default arguments are:
\begin{footnotesize}
\begin{verbatim}
npr.proc_lv3(flat_mask=0.3, flat_fill=0.3)
\end{verbatim}
\end{footnotesize}

This is a standard process of dark subtraction and flat-fielding. In NIC, the flat image near the edge of each lit part has a very small flat value. Thus, those edge region will have very high pixel values in the processed image. 
Because of this, by default, it fixes flat frame values $ < 0.3 $ (\texttt{flat\_mask}) to a constant value $ 0.3 $ (\texttt{flat\_fill}) before flat-fielding. From level 3, each FITS frame is split into two FITS files, for o-ray and e-ray, specified by a suffix in the file name (Sect. \ref{ss: fits name}). This greatly reduces the size of each image (from 4.2 MB to two files of 0.28 MB). Finally, using \texttt{mmask} for the corresponding filter (detector), bad pixel interpolation is conducted. It prioritizes the vertical direction over the horizontal direction when the masked region has an identical length to both directions (so that the result is closer to \texttt{IRAF.PROTO.FIXPIX}). After processing, the summary file \texttt{summ(lv3)} is made in the log directory.

\vspace{1em}
\subsubsection{Cell 7: lv4 Planner for Fringe Frames}\label{sss: cell7}
The important default arguments are:
\begin{footnotesize}
\begin{verbatim}
npr.plan_ifrin(add_crrej=True)
\end{verbatim}
\end{footnotesize}

\begin{figure} [tb!]
  \begin{center}
    \includegraphics[width=\linewidth]{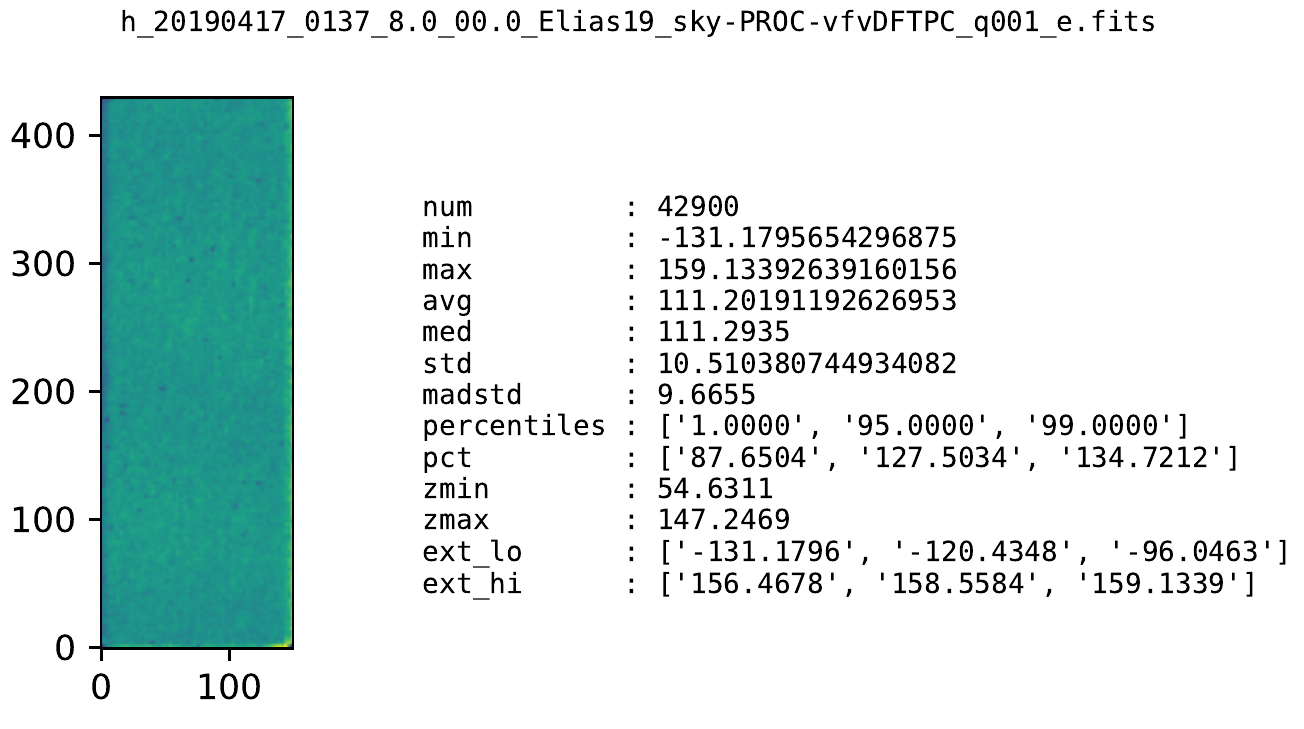}
  \end{center}
  \caption{An example of saved thumbnail images in \texttt{\_\_logs/<name>/thumbs\_mfrin\_plan/}. From the title, this is an H-band e-ray blank sky frame, taken nearby the object Elias19, at \texttt{POL-AGL1 = 0.0}. The statistics on the right part are identical to Fig. \ref{fig:mdarkplan}.}
  \label{fig:mfrinplanthumb}
\end{figure}

The fringe frames are selected by finding any frame that has header \texttt{OBJECT} value ends with \texttt{\_sky}. For fringe frames, the thumbnails are saved into \texttt{thumbs\_mfrin\_plan/} in the log directory (Fig. \ref{fig:mfrinplanthumb}). Also, the the planner \texttt{plan(IFRIN)} is made in the log directory. In the planner, there are columns for parameters of \texttt{astroscrappy} (cosmic-ray rejection) for each image. Currently, Gaussian or Moffat convolution options for fine structure calculation are not implemented. They are not important, as the image is intended to be a blank sky frame without any source. Among the parameters, \texttt{sepmed=False} makes \texttt{astroscrappy} runs similar to the original L.A.Cosmic, and we found this is essential for properly removing cosmic-rays on NIC. Then \texttt{sigclip=4.5, sigfrac=5, objlim=1, satlevel=30000, cleantype="medmask", fs="median"} are used by default, with proper gain and readout noise values for each detector.

\vspace{1em}
\subsubsection{Cell 8: Make lv4 Fringe \& Master Fringe Planner}\label{sss: cell8}
There are no important arguments the user can tune. It makes level 4 image of fringe frames based on \texttt{plan(IFRIN)}, saves them into \texttt{frin/} sub-directory inside the level 4 directory, and makes the next planner file \texttt{plan(MFRIN)}. In \texttt{plan(MFRIN)}, the most important column is \texttt{FRINCID}, which stands for fringe combination identifier. Any frames with the same \texttt{FRINCID} will be combined into one single master fringe frame. By default, it is set \texttt{<FILTER>\_<OBJECT>\_<id>\_<OERAY>}. Except for \texttt{<id>}, they are as in Sect. \ref{ss: fits name}. \texttt{<id>} is an integer that separates blank sky exposures that are separated by another exposure. For example, blank sky exposures before and after the exposures of the main target will be separated by \texttt{<id>} of \texttt{001} and \texttt{002} automatically. By default, \texttt{POL-AGL1} The user can put non-zero value in \texttt{REMOVEIT} to ignore it for master fringe combination.

\vspace{1em}
\subsubsection{Cell 9: Make Master Fringe}\label{sss: cell9}
The important default arguments are:
\begin{footnotesize}
\begin{verbatim}
npr.comb_mfrin(
    combine_kw=dict(
      combine="med",
      reject="mm",
      n_minmax=(0, 1),
      scale="avg_sc",
      scale_section="[20:100, 20:400]",
      scale_to_0th=False,
    ),
)
\end{verbatim}
\end{footnotesize}

Based on the planner, master fringe is made by combining the same-\texttt{FRINCID} frames. By default, each fringe frame is divided by sigma-clipped average value (\texttt{scale}) within \texttt{scale\_section}, so that the sky level variation is removed. By default, in \texttt{IRAF.IMMATCH.IMCOMBINE}, ``scale factors are normalized so that the first input image has no scaling'', which is reproducible by \texttt{scale\_to\_0th=True}\footnote{
  Note that the results may be different from the IRAF result because IRAF calculates statistics by selecting only a few pixels from the region to reduce computation time and memory usage, and this behavior is hard to modify.
}. This is useful when Poisson noise calculation is important. By default, \texttt{NICpolpy} scales each image before combination. \texttt{reject="mm", n\_minmax=(0,1)} means it rejects 0 smallest and 1 largest pixel values, before median (\texttt{combine="med"}) combination. 

\begin{figure} [tb!]
  \begin{center}
    \includegraphics[width=\linewidth]{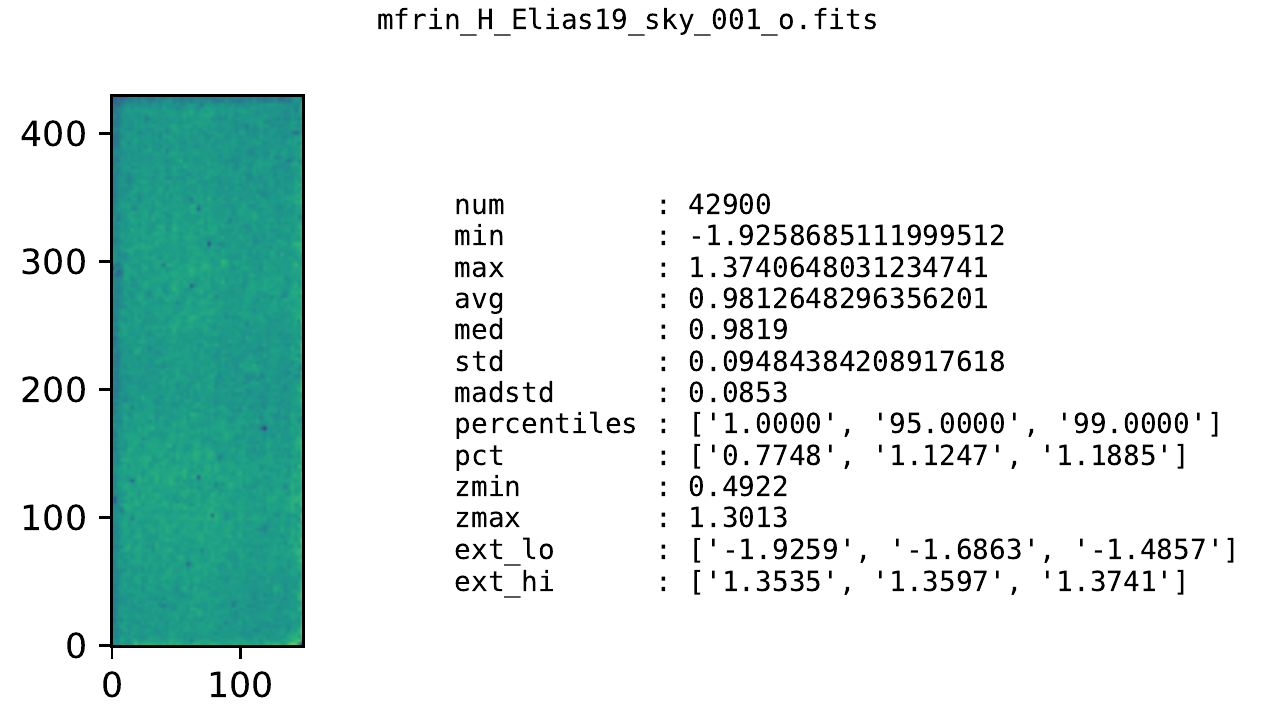}
  \end{center}
  \caption{An example of saved thumbnail images in \texttt{\_\_logs/<name>/thumbs\_mfrin/}. This is a result of combining four fringe frames, including that in Fig. \ref{fig:mfrinplanthumb}. The statistics on the right part are identical to Fig. \ref{fig:mdarkplan}.}
  \label{fig:mfrinthumb}
\end{figure}

The combined frames are saved in the log directory (\texttt{cal-mfrin/}) with thumbnails (\texttt{thumbs\_mfrin/}) and summary file \texttt{summ(MFRIN)}. An example of a combined master fringe thumbnail is shown in Fig. \ref{fig:mfrinthumb}.

\vspace{1em}
\subsubsection{Cell 10: lv4 Planner}\label{sss: cell10}
The important default arguments are:
\begin{footnotesize}
\begin{verbatim}
npr.plan_lv4(add_crrej=True, add_mfrin=True)
\end{verbatim}
\end{footnotesize}

The final level 4 processing is cosmic-ray removal and fringe subtraction. Those can be turned on or off by \texttt{add\_crrej} and \texttt{add\_mfrin}. Here, we used \texttt{add\_mfrin=False}. The planner \texttt{plan(lv4)} is made, and it contains multiple columns related to the cosmic-ray rejection algorithm as in \texttt{plan(IFRIN)} (see Sect. \ref{sss: cell7}). If \texttt{add\_mfrin=True}, a column named \texttt{FRINFRM} is added to indicate the corresponding fringe frame.

\vspace{1em}
\subsubsection{Cell 11: Make lv4}\label{sss: cell11}
The important default arguments are:
\begin{footnotesize}
\begin{verbatim}
npr.proc_lv4(
    frin_bezels=20,
    frin_sep_kw=dict(minarea=np.pi*5**2),
    frin_scale_kw=dict(sigma=2.5),
)
\end{verbatim}
\end{footnotesize}    

The important arguments are the parameters related to fringe subtraction. First, \texttt{NICpolpy} loads the corresponding fringe frame based on \texttt{FRINFRM} in \texttt{Plan(lv4)}. Then it crops the 20-pixel edges (\texttt{frin\_bezels}). Within the central region, it finds any extended bright object using SExtractor algorithm \citep{1996A&AS..117..393B}. The variance map is calculated by proper gain and readout noise parameters, and any object above the 5-sigma level and segment area of equivalent circular radius of 5 pixels (\texttt{frin\_sep\_kw}) is found. The fringe frame is then scaled based on these pixels (sigma-clipped median is used; \texttt{frin\_scale\_kw}) and subtracted from the science exposure image. Finally, cosmic-rays are removed based on the parameters specified in \texttt{plan(lv4)}.

\vspace{1em}
\subsubsection{FITS Header}\label{sss: header}
Finally, each FITS file contains much information on how it is processed throughout \texttt{NICpolpy}. A few keywords are inserted:
\begin{itemize}
  \item \texttt{[MIN/MAX]V<iii>[O/E]} (e.g., \texttt{MINV001E}): \texttt{<iii>}-th minimum or maximum pixel values in level 1 image in o-/e-ray region.
  \item \texttt{NSATPIX[O/E]}: The number of saturated pixels in o-/e-ray region, based on level 1 image (Sect. \ref{sss: cell4}).
  \item \texttt{FFTCUTWL}: The minimum wavelength of Fourier pattern removal used for level 2 (Sect. \ref{sss: cell5})
  \item \texttt{LViFRM}: The path to the level \texttt{i} frame related to this file.
  \item \texttt{[DARK/FLAT/FRIN]FRM}: The master dark, flat, or fringe frame used for this file.
  \item \texttt{MASKFILE}: The path to \texttt{mmask} file.
  \item \texttt{MASKNPIX}: The number of masked (fixed) pixels in level 3 (Sect. \ref{sss: cell6}).
  \item \texttt{CRNFIX}: The number of pixels fixed by cosmic-ray rejection algorithm in level 4 (Sect. \ref{sss: cell11}).
  \item others: \texttt{YYYYMMDD}, \texttt{COUNTER}, \texttt{SETID}, \texttt{OERAY}, \texttt{PROCESS} as in Sect. \ref{ss: fits name}.
\end{itemize} 

Also, as \texttt{COMMENT} and \texttt{HISTORY}, most of the minor logs for this specific file are saved, with options used, timestamp, and the time taken for the step. Below is the last part of the header of a level 4 image:
\begin{tiny}
\begin{verbatim}
HISTORY ================== Level 1 (vertical pattern removal) ==================
HISTORY Extrema pixel values found N(smallest, largest) = (5, 5) excluding mask 
HISTORY (__logs/SP_20190417/cal-mmask/mmask_H.fits) and bezel: ((20, 20), (20, 2
HISTORY 0)) in xyz order. See MINViii[OE] and MAXViii[OE].                      
HISTORY Saturated pixels found based on satlevel = 8000, excluding mask (__logs/
HISTORY SP_20190417/cal-mmask/mmask_H.fits) and bezel: ((20, 20), (20, 20)) in x
HISTORY yz order. See NSATPIX and SATLEVEL.                                     
HISTORY ..................................(dt = 0.009 s) 2022-09-25T09:32:30.900
HISTORY Vertical pattern subtracted using ['[:, 100:250]', '[:, 850:974]'] by ta
HISTORY king median with sigma-clipping in astropy (v 5.1), given {'sigma': 2, '
HISTORY maxiters': 5}. (using pixel mask) maskpath='__logs/SP_20190417/cal-mmask
HISTORY /mmask_H.fits'                                                          
HISTORY ..................................(dt = 0.019 s) 2022-09-25T09:32:30.920
COMMENT [yfu.update_process] Standard items for PROCESS includes B=bias, D=dark,
COMMENT  F=flat, T=trim, W=WCS, O=Overscan, I=Illumination, C=CRrej, R=fringe, P
COMMENT =fixpix, X=crosstalk.                                                   
COMMENT User added items to PROCESS: v=vertical pattern.                        
HISTORY ------------------------------------------------------------------------
HISTORY ================== Level 2 (Fourier pattern removal) ===================
HISTORY FIXPIX on the left part of the image                                    
HISTORY ..................................(dt = 0.512 s) 2022-09-25T09:34:56.037
HISTORY Median filtered (convolved) frame calculated with {'size': 5, 'footprint
HISTORY ': None, 'mode': 'reflect', 'cval': 0.0, 'origin': 0}                   
HISTORY ..................................(dt = 0.168 s) 2022-09-25T09:34:56.207
HISTORY Sky standard deviation (MB_SSKY) calculated by sigma clipping at MB_SSEC
HISTORY T with {'sigma': 3.0, 'maxiters': 50, 'std_ddof': 1}; used for std_ratio
HISTORY  map calculation.                                                       
HISTORY ..................................(dt = 0.020 s) 2022-09-25T09:34:56.228
HISTORY [medfilt_bpm] Median-filter based Bad-Pixel Masking (MBPM) applied.     
HISTORY ..................................(dt = 0.004 s) 2022-09-25T09:34:56.232
HISTORY FFT(left half) to get pattern map (see FFTCUTWL for the cut wavelength);
HISTORY  subtracted from both left/right.                                       
HISTORY ..................................(dt = 0.002 s) 2022-09-25T09:34:56.241
COMMENT User added items to PROCESS: f=fourier pattern.                         
HISTORY Vertical pattern subtracted using ['[:, 100:250]', '[:, 850:974]'] by ta
HISTORY king median with sigma-clipping in astropy (v 5.1), given {'sigma': 2, '
HISTORY maxiters': 5}. (using pixel mask) maskpath='__logs/SP_20190417/cal-mmask
HISTORY /mmask_H.fits'                                                          
HISTORY ..................................(dt = 0.019 s) 2022-09-25T09:34:56.263
HISTORY ------------------------------------------------------------------------
HISTORY [yfu.darkcor] Dark subtracted (DARKFRM = __logs/SP_20190417/cal-mdark/md
HISTORY ark_H_8.0s.fits)                                                        
HISTORY ..................................(dt = 0.001 s) 2022-09-25T09:38:34.343
HISTORY [yfu.flatcor] Flat pixels with `value < flat_mask = 0.3` are replaced by
HISTORY  `flat_fill = 0.3`.                                                     
HISTORY .................................................2022-09-25T09:38:34.345
HISTORY [yfu.flatcor] Flat corrected (FLATFRM = __logs/SP_20190417/cal-mflat/mfl
HISTORY at_H_20180507-lv1.fits)                                                 
HISTORY ..................................(dt = 0.002 s) 2022-09-25T09:38:34.346
HISTORY [yfu.imslice] Sliced using `trimsec = '[741:890, 331:760]'`: converted t
HISTORY o (slice(330, 760, None), slice(740, 890, None)).                       
HISTORY ..................................(dt = 0.001 s) 2022-09-25T09:38:34.352
HISTORY [fixpix] Pixel values interpolated.                                     
HISTORY ..................................(dt = 0.191 s) 2022-09-25T09:38:34.736
HISTORY Cosmic-Ray rejection (CRNFIX=0 pixels fixed) by astroscrappy (v 1.1.1.de
HISTORY v8+g783f217). Parameters: {'gain': 9.8, 'readnoise': 36, 'sigclip': 4.5,
HISTORY  'sigfrac': 5.0, 'objlim': 1, 'satlevel': 30000, 'niter': 4, 'sepmed': F
HISTORY alse, 'cleantype': 'medmask', 'fsmode': 'median'}                       
HISTORY ..................................(dt = 0.041 s) 2022-09-25T09:42:40.204
\end{verbatim}
\end{tiny}

\vspace{1em}
\section{Polarimetric Dome Flat Frames} \label{s: flat}
A large number of polarimetric dome flat frames were obtained on UT 2020-06-03 (total 640 frames per filter; 8 \texttt{INSROT} and 4 \texttt{POL-AGL1} combinations and 20 FITS frames per each combination). These frames are useful to check if there is any region in the RoI the observer should avoid putting the scientific targets. Also, these flat frames were used to re-calculate the gain factor (Sect. \ref{s: gain rdn}).

The frames underwent these processes: 
\begin{enumerate}
  \item Reduce all frames to level 1.
  \item Split o-/e-ray regions by cropping only the lit parts (Fig. \ref{fig:fignicfitsanatomy}).
  \item Scale each frame by ``median value/median value of the first frame''.
\end{enumerate} 

The master flat frame is obtained by considering all 640 frames per filter. The flat frames were also grouped by different combinations: 20 frames per each \texttt{(INSROT, POL-AGL1)} pair, 80 frames per each \texttt{INSROT}, and 160 frames per each \texttt{POL-AGL1}, which result in 32, 8, and 4 groups per filter, respectively. The median and sample standard deviation of each pixel after 3-sigma clipping are saved. The signal-to-noise (SNR) map is generated by dividing the median map by the sample standard deviation map. 

\begin{figure} [tb!]
  \begin{center}
    \includegraphics[width=\linewidth]{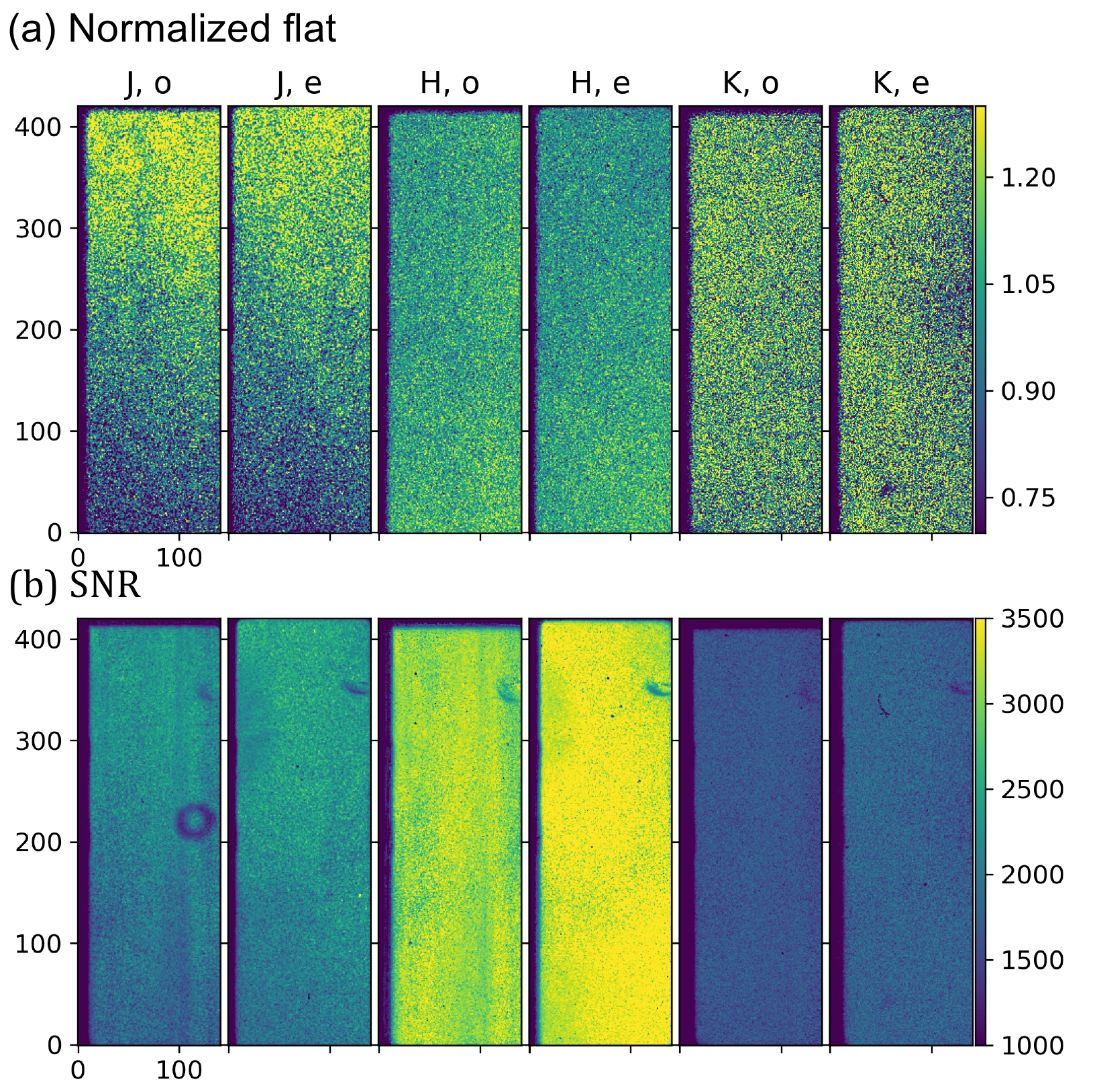}
  \end{center}
  \caption{The master flat frames for UT 2020-06-03 data. The top row (a) is the normalized flat, and the bottom row (b) is the signal-to-noise ratio distribution.}
  \label{fig:fsnr}
\end{figure}

The master flats (normalized) and SNR maps are shown in Fig. \ref{fig:fsnr}. Two interesting features are found: A common doughnut-shaped region at the top-right corner in SNR maps and a large doughnut-shaped region in the J-band o-ray region in the mid-right part. These are visible only in the SNR map but not in the master flat map. These doughnuts could also be found in the flat frames taken on UT 2018-05-07 using a similar technique.

\begin{figure*} [tb!]
  \begin{center}
    \includegraphics[width=\linewidth]{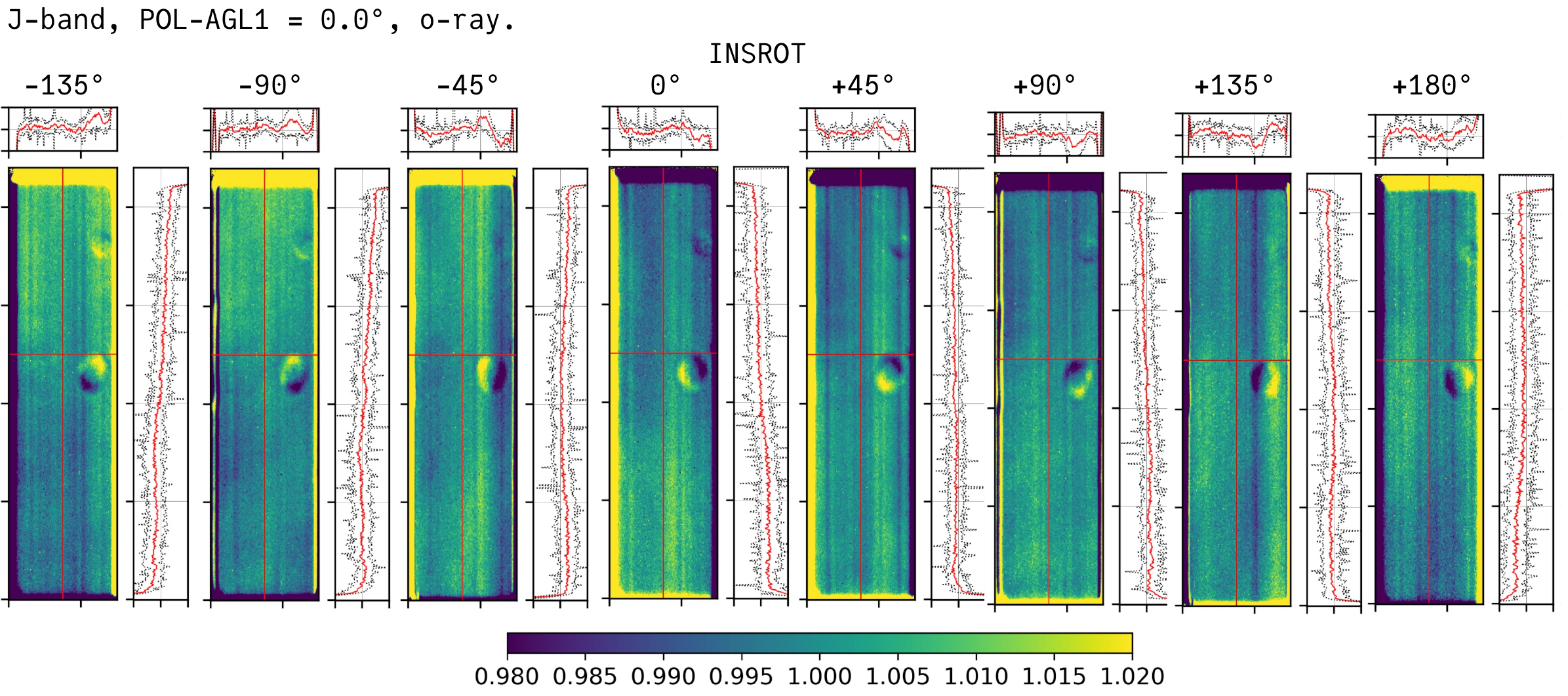}
  \end{center}
  \caption{The $ r_1 $ ratio map for J-band with HWP rotator angle $ 0^\circ $ and o-ray region only. The numbers on top are the \texttt{INSROT} values of the group. The vertical and horizontal graphs show the cut profiles along red lines in the flat images. The red lines in the graphs are the cut profile of $ \pm 10 \,\mathrm{pixel} $ around the red lines in the image, and black dotted lines are the minimum and maximum pixel values within the $ \pm 10 \,\mathrm{pixel} $ region. The ticks for the images indicate 100 pixels, while the ticks for the cut prociles are $ 0.98 $, $1.00$, and $1.02$.}
  \label{fig:flatinsrot}
\end{figure*}

An interesting feature appeared for flats grouped based on \texttt{INSROT}. Define the ratio map as
\begin{equation}
  r_1(\texttt{INSROT}, \texttt{POL-AGL1}) 
    = \frac{\mathrm{comb}(\texttt{INSROT}, \texttt{POL-AGL1})}
      {\mathrm{comb}(\mathrm{any}, \texttt{POL-AGL1})}
\end{equation}
for the given \texttt{(INSROT, POL-AGL1)} pair, i.e., the combined image for the given \texttt{(INSROT, POL-AGL1)} pair divided by the image combining all frames for the same \texttt{POL-AGL1}, to see the effect of \texttt{INSROT}. An example for J-band o-ray region for \texttt{POL-AGL1 = 0} are shown in Fig. \ref{fig:flatinsrot}. The relative location of the brighter and darker part of the common doughnut \textit{rotates} with increasing \texttt{INSROT}. Hence, it is likely that dust particles or defects in the common optical path are responsible for the common doughnuts (top-right corner). The master flat may be different from the true flat at the observation by up to $ \sim 2\mathrm{-}3 \% $ at such regions, depending on the \texttt{INSROT} at the exposure. For safety, any observer is encouraged to put their scientific object such that they are not affected by these doughnuts. Also, the vertical cut profile has a gradient of $ \sim \pm 1\% $, changing as a function of \texttt{INSROT}. The change in the lit part is seen, too. 

A similar ratio map was made to investigate the effect of \texttt{POL-AGL1}:
\begin{equation}
  r_2(\texttt{INSROT}, \texttt{POL-AGL1}) 
  = \frac{\mathrm{comb}(\texttt{INSROT}, \texttt{POL-AGL1})}
  {\mathrm{comb}(\texttt{INSROT}, \mathrm{any})}~.
\end{equation}
In $ r_2 $ maps, the dust doughnuts disappear. This is likely because the dark-bright features of doughnuts cancel out when combined over the \texttt{INSROT}. The combined flats show some large-scale pattern in the final flat, with fluctuation $ \lesssim \pm 0.5\% $ level for all three bands, except for the edge regions. 

To summarize, the users are recommended to avoid the top-right and middle-right regions of the RoI to best use the three-band data. Also, considering the bad pixel mask based on both the dark and flat (Fig. \ref{fig:imask}), it is better to avoid the top-left part due to the over density of bad pixels in H-band (e-ray) and K-band (e-ray). These regions may change over time as new defects appear.

The results for different cases (figures and videos for both $ r_1 $ and $ r_2 $) are available in the supporting material in Appendix \ref{s: sm}.

\section{Gain and Readout Noise: Revisited} \label{s: gain rdn}
The gain (the conversion factor) and readout noise are important parameters in the error analysis and data reduction (e.g., used in fringe frame scaling, cosmic-ray rejection, and object detection algorithms). They were first reported in \citet{Ishiguro2011ARNHAO}. According to them, the expected noise in the bias frame corresponds to the readout noise divided by the gain factor, i.e., 5.4, 7.7, and 8.8 ADUs for J, H, and K-band, respectively. However, the sample standard deviations in the dark area in level 1 images are smaller (roughly 4 ADU for all detectors). This was the motivation for re-calculating the two detector parameters. We also briefly tested if there is any hint that these parameters are different per pixel.

If the two flat images have pixel values of $ F_1 $ and $ F_2 $ and two biases have $ B_1 $ and $ B_2 $ (all in ADU), the gain and readout noise are
\begin{equation}\label{eq: janesick g}
  g = \frac{ (\bar{F}_1 + \bar{F}_2) - (\bar{B}_1 + \bar{B}_2)}{\sigma^2_{F_1 - F_2} - \sigma^2_{B_1 - B_2}} ~\mathrm{[e/ADU]}
\end{equation}
and
\begin{equation}\label{eq: janesick r}
    R = g\frac{\sigma_{B_1 - B_2}}{\sqrt{2}} ~\mathrm{[e]} ~.
\end{equation}
Here $ \bar{X} $ means the average of all the pixels in the frame $ X $, and $ \sigma_X $ is the true standard deviation of the frame $ X $, estimated from the sample standard deviation $ \sigma_X \approx [(\sum_i (X_i - \bar{X})^2) / (N - 1)]^{1/2} $. This method is called Janesick's method. The proof is given in Appendix \ref{s: Janesick}.

In the case of NIC, the bias is effectively 0, and the standard deviation of the bias-difference map is obtained by the difference map of short-exposure darks to minimize the Poisson noise effect. We selected 20 level 1 dark frames with 2 s exposure time taken on UT 2019-10-22 for each of the three filters (detectors) to calculate $ R/g $. Next, the flat frames taken on UT 2020-06-04 are used for $ g $ calculation so that both $ g $ and $ R $ can be obtained. From each dark and flat frame, the bad pixels (\texttt{imask}, Sect. \ref{s: lv1}) are masked.

\vspace{1em}
\subsection{$ R/g $} \label{ss: rdn}
Here, we discuss how we tested the pixel-to-pixel and quadrant-to-quadrant variabilities of the $ R/g $ value. Then the absolute values were obtained in ADU for each detector. 

\vspace{1em}
\subsubsection{Pixel-to-Pixel Variability}
\begin{figure} [tb!]
  \begin{center}
    \includegraphics[width=\linewidth]{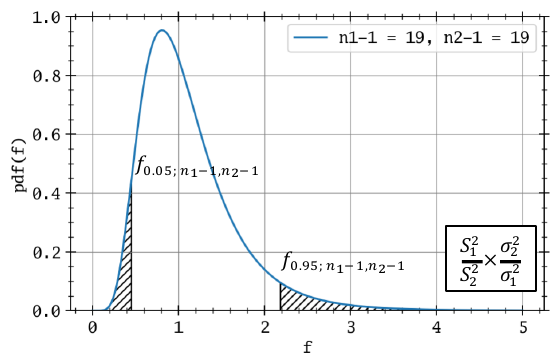}
  \end{center}
  \caption{An example of F-distribution probability distribution function (pdf) with degrees of freedom $ (n_1-1, n_2-1) = (19, 19) $. The rejection regions of significance level $ \alpha = 0.10 $ for a two-tailed test are shown by two hatched regions. }
  \label{fig:fdistn}
\end{figure}

Are the $ R/g $ values dependent on each pixel, i.e., each pixel has different readout noise per gain? Say the sample standard deviation of the pixel at \texttt{(i,j)} among the 20 dark frames is $ S^{ij} $ and the true standard deviation is $ \sigma^{ij} $ (both in ADU). Since the dark frames have a very short exposure time and most hot pixels are masked, the only dominant source of the scatter is readout noise or $ R/g $. If the hypotheses are set as
\begin{equation}\label{eq: dk hyp}
  H_0: \sigma^{ij} = \sigma^{i'j'} 
  \quad;\quad 
  H_1: \sigma^{ij} \neq \sigma^{i'j'} ~,
\end{equation}
we can test the null hypothesis ($ H_0 $) using the F-statistic (e.g., \cite{walpole}, Sect. 4.5). Defining
\begin{equation}\label{eq: F-stat}
  F := \frac{S_1^2}{S_2^2} \times \frac{\sigma_2^2}{\sigma_1^2} \sim F_{n_1-1, n_2-1} ~,
\end{equation}
where $ \sim $ means the statistic in the left-hand side follows the distribution in the right-hand side, and $ F_{n_1-1, n_2-1} $ means the F-distribution of degrees of freedom $ (n_1-1, n_2-1) $. $ n_1 $ and $ n_2 $ are the number of sample used to obtain $ S_1 $ and $ S_2 $, respectively. Under $ H_0 $, the value of the statistic $ f = S_1^2 / S_2^2 $. Fig. \ref{fig:fdistn} visually depicts how the rejection process works: If our calculated $ f $ value is inside one of those hatched regions, our null hypothesis is rejected with significance level $ \alpha = 0.10 $, i.e., $ \sigma_1 \neq \sigma_2 $. If our statistic is not in the hatched regions, we can make no conclusion under the significance level $ \alpha = 0.10 $. A number that is more frequently used is the P-value: It measures the degree of rarity of the observation ($ f $ value) under $ H_0 $. Consider two cases where one obtains $ f = f_{0.20 ; n_1-1, n_2-1} $ and $ f = f_{0.02 ; n_1-1, n_2-1} $. the P-value will be 0.4, and 0.04, respectively, which means the latter case is rarer than the former one. Thus, the null hypothesis is rejected in the latter case ($ H_0 $ is unlikely to be true because such a rare observation was made), while it is not in the former case.

\begin{figure} [tb!]
  \begin{center}
    \includegraphics[width=\linewidth]{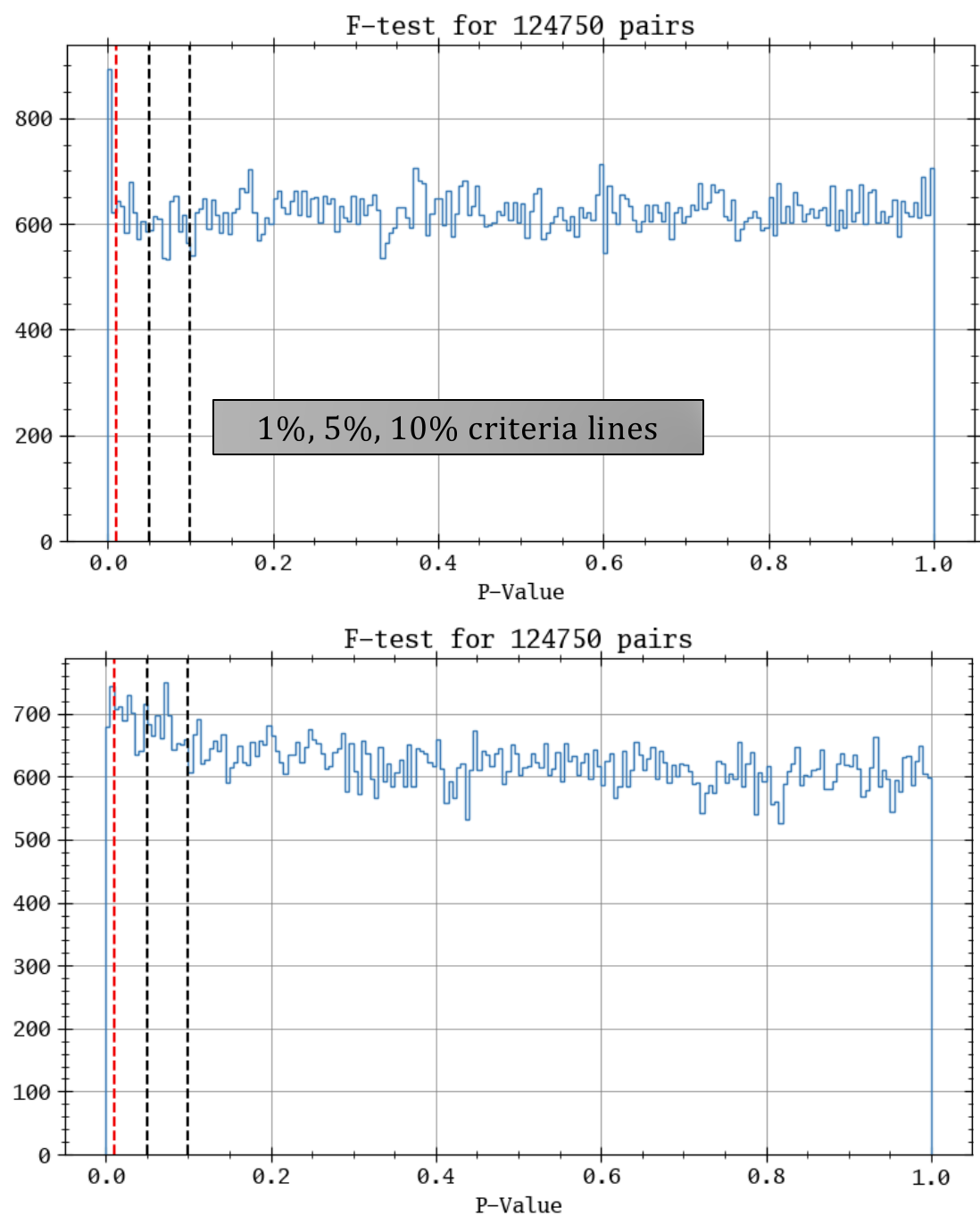}
  \end{center}
  \caption{Two examples of P-value calculations and their distributions for $ 124,750 $ random pixel pairs. The settings for obtaining these two figures are identical, except these two simulations samples different random pixel pairs. The three vertical lines from left to right mean significance s of 0.01, 0.05, and 0.10, respectively.}
  \label{fig:ftest}
\end{figure}

Because we have $ \sim 10^6 $ pixels at each filter, we have $ \sim 10^{12} $ possible F-statistic values (pairs). To save time, we randomly selected $ \sim 10^5 $ pairs, calculated the P-value for each pair, and saw the distribution of the P-values. Fig. \ref{fig:ftest} shows two trials of such test on K-band detector dark frames. We found no significant population of the histogram at small P-values, even at multiple trials and different detectors. This means we cannot reject the null hypothesis in the vast majority of cases. Therefore, we have no clear evidence to conclude that $ R/g $ is pixel-dependent.

\vspace{1em}
\subsubsection{$ R/g $ Value}
\begin{figure} [tb!]
  \begin{center}
    \includegraphics[width=\linewidth]{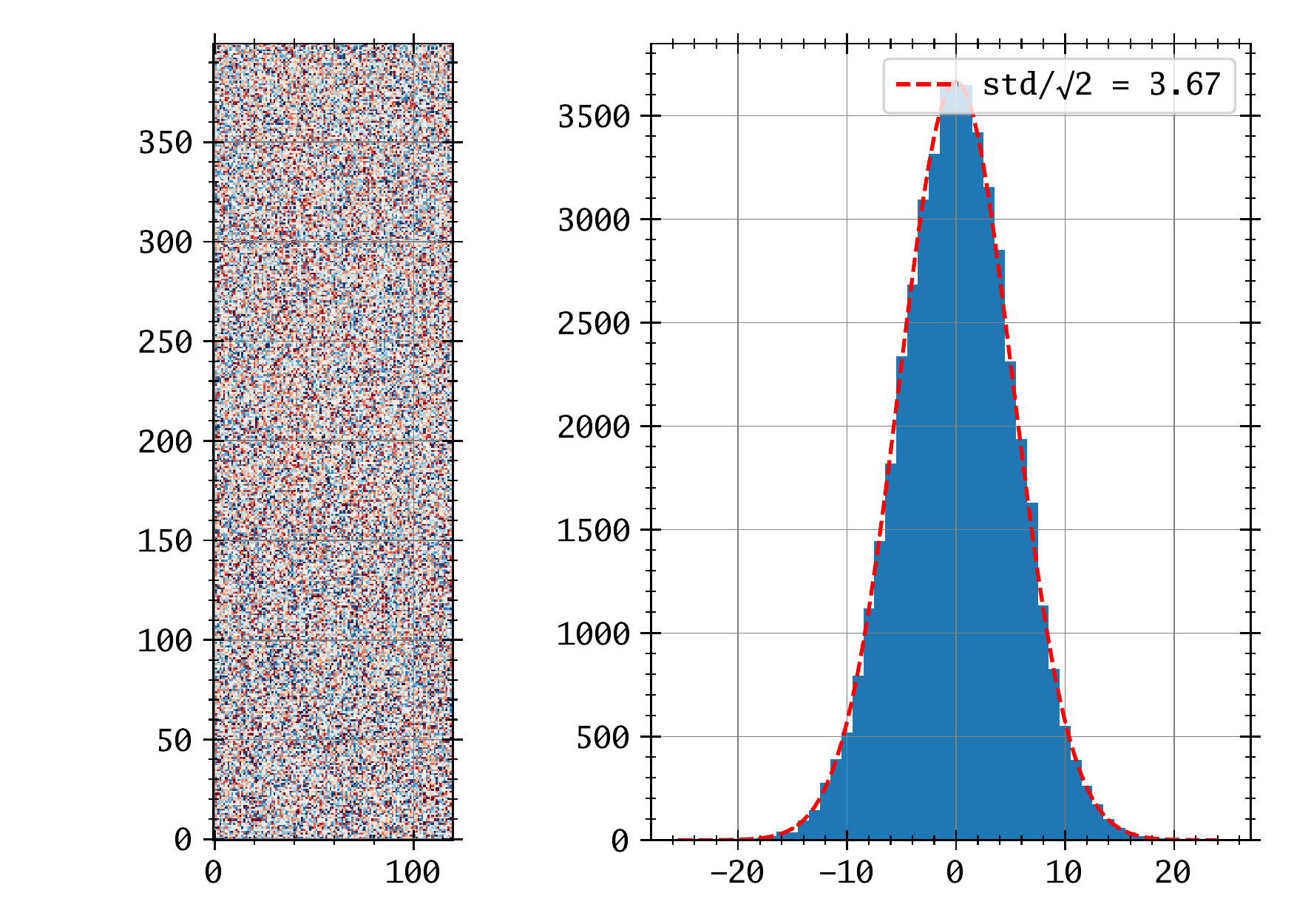}
  \end{center}
  \caption{The difference between two random K-band dark frames (left; blue and red are $ -10 $ and $ +10 $ ADU, respectively), and the pixel value distribution histogram (right). The fitted Gaussian function is shown in the red dashed line with its standard deviation information.}
  \label{fig:darksub}
\end{figure}

As a quick test, we selected two random dark frames in K-band, cropped only the o-ray lit part, and pixels that have values outside the 5-sigma for each frame are masked. These two dark frames are subtracted, as in Fig. \ref{fig:darksub}. The histogram shows the distribution is very close to a normal distribution as one would expect, and the standard deviation divided by $ \sqrt{2} $, i.e., $ R/g = 3.67 \,\mathrm{ADU} $ for this specific case.

\begin{figure} [tb!]
  \begin{center}
    \includegraphics[width=\linewidth]{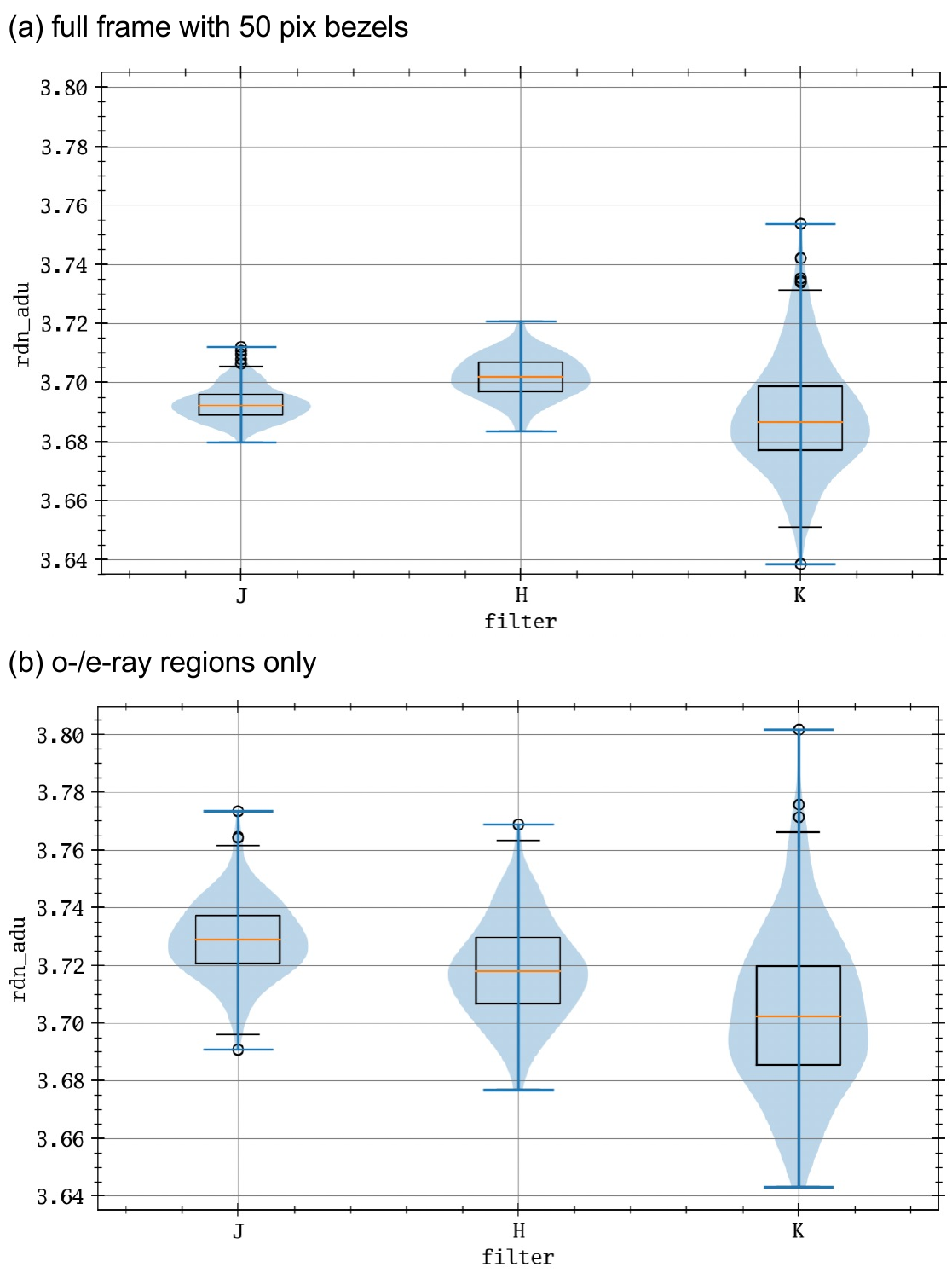}
  \end{center}
  \caption{The distribution of measured $ R/g $ for three detectors. The measurement used all pixels within the 50-pixel edge (a) and only the o-/e-ray lit parts (b). The box-whisker plots with median are shown, and the violin plot is overplotted. The boxes indicate the first ($ Q_1 $) and third ($ Q_3 $) quantiles, and the black whiskers extend until $ \min\{ \min\{y\}, Q_1 - 1.5 (Q_3-Q_1)\} $ to the bottom and $ \max\{ \max\{y\}, Q_3 + 1.5 (Q_3-Q_1)\} $ to the top, where $ \min\{y\} $ and $ \max\{y\} $ are the minimum and maximum of the data, respectively. Black circles show the outliers (fliers) outside the whisker.}
  \label{fig:rdnall}
\end{figure}

Since there are total 20 dark frames, one can make $ \frac{n(n-1)}{2} = 190 $ pairs of dark frames and see how the $ R/g $ is distributed. The result is shown in Fig. \ref{fig:rdnall}. Regardless of the regions we chosen, the final $ R/g $ is well constrained to $ \sim 3.7 \, \mathrm{ADU} $ for all three detectors, with scatter much less than $ 0.1 \,\mathrm{ADU} $. 

\vspace{1em}
\subsubsection{Quadrant-to-Quadrant Variability}
\begin{figure*} [tb!]
  \begin{center}
    \includegraphics[width=\linewidth]{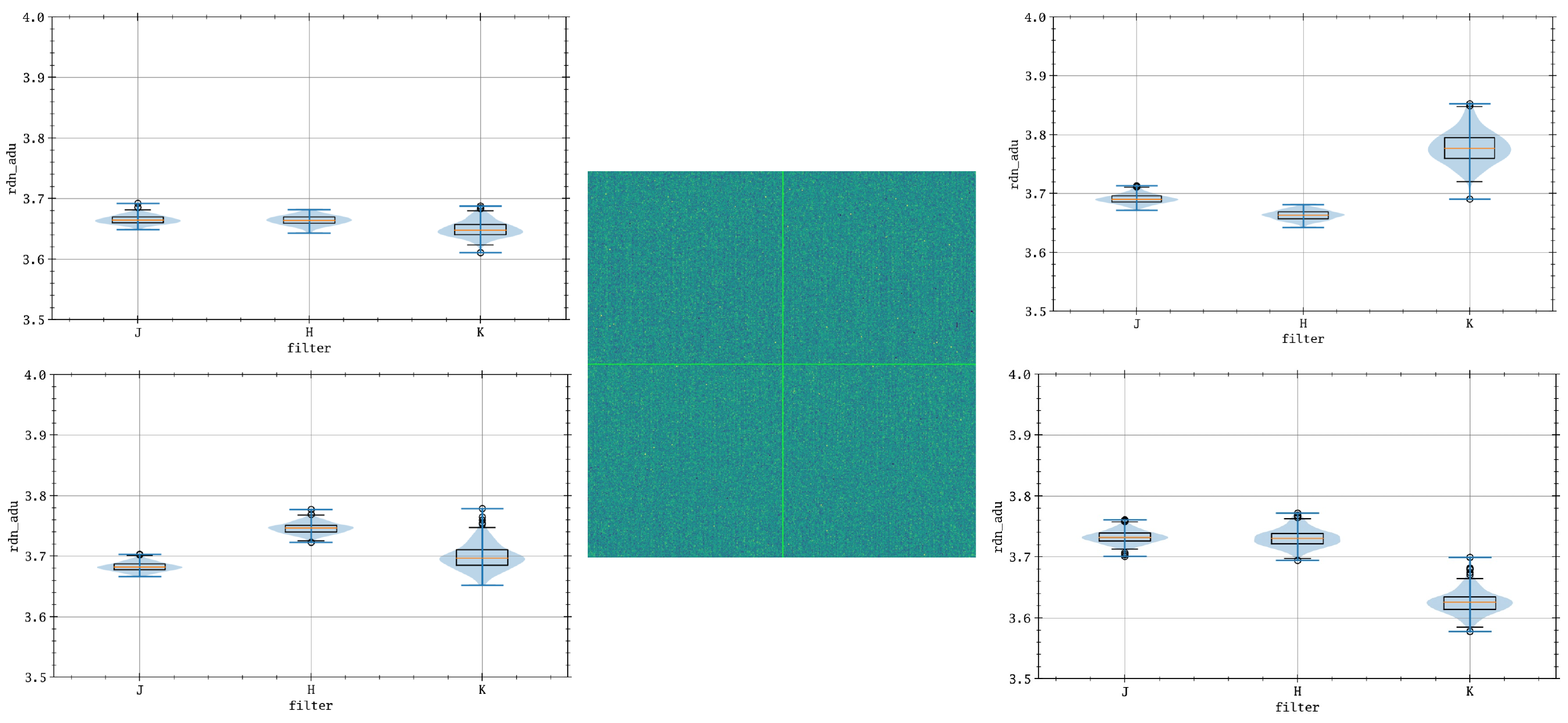}
  \end{center}
  \caption{The distribution of measured $ R/g $ for three detectors at each quadrant. The plots in the upper left, upper right, lower left, and lower right show the $ R/g $ measurement distribution of 190 pairs for the upper left, upper right, lower left, and lower right quadrants, respectively.}
  \label{fig:rdnquad}
\end{figure*}

Finally, we tested if the $ R/g $ depends on the quadrants. The technique, the same as Fig. \ref{fig:rdnall}, is applied to each quadrant and checked if any clear difference could be seen (Fig. \ref{fig:rdnquad}). Although the distribution is different for quadrants in K-band, the difference is only $ \sim 0.1 \,\mathrm{ADU} $. Therefore, we conclude there is no quadrant dependency in $ R/g $.

\vspace{1em}
\subsection{Gain (The Conversion Factor) \& Readout Noise}
Hinted by the $ R/g $ calculation, we now assume both the gain and readout noise are constant values for all pixels in each detector. As before, NIC has no bias, so $ \bar{B} = 0 $. Since $ R/g $ is a constant, $ \sigma^2_{B_1 - B_2} $ in Eq. (\ref{eq: janesick g}) can be replaced by $ 2(R/g)^2 $, and
\begin{equation}
  F^{i j} \sim \mathcal{N}\left(f^{i j}, \frac{f^{i j}}{g} + 2\left( \frac{R}{g} \right)^2 \right) ~,
\end{equation}
for the flat frame pixel value variable $ F^{i j} $ and its measured value $ f^{i j} $ at the pixel \texttt{(i,j)}. Here, $ \mathcal{N}(\mu, \sigma^2) $ is the normal distribution with mean $ \mu $ and variance $ \sigma^2 $. The two terms in the variance are the Poisson and readout noise terms. Say the true pixel value (i.e., flux) of $ F^{i j} $ is constant. Then one can calculate the average of the location $ \bar{F}^{ij} $ and sample standard deviation $ S^{ij} $ to obtain
\begin{equation} \label{eq: gain calc}
  g = \frac{\bar{F}^{ij}}{S^{ij} - (R/g)^2} ~.
\end{equation}
Here, $ R/g = 3.7 \,\mathrm{ADU} $ is already obtained previously and found to be constant. 

\begin{figure*} [tb!]
  \begin{center}
    \includegraphics[width=0.85\linewidth]{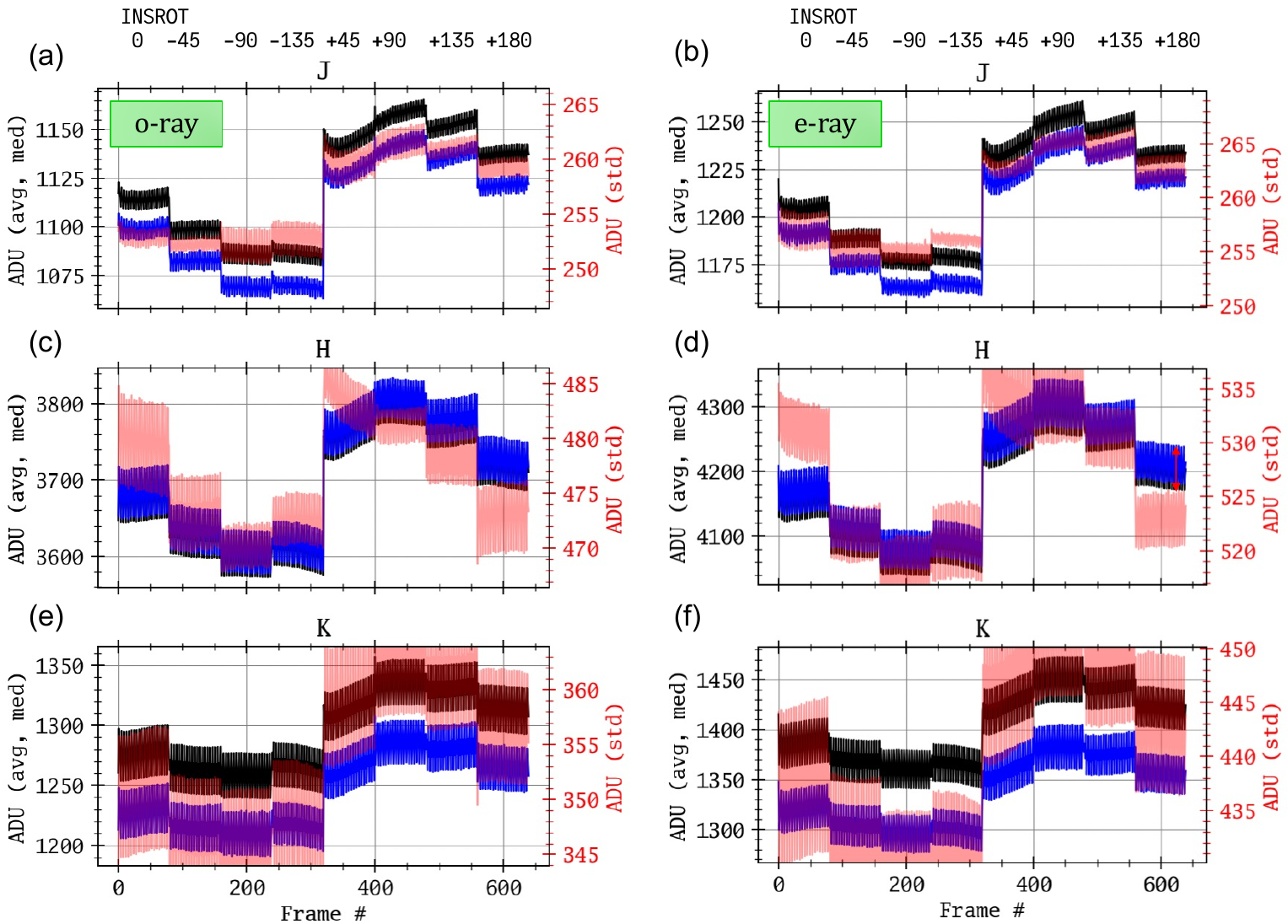}
  \end{center}
  \caption{The variability of flat counts in J-band (a, b), H-band (c, d), and K-band (e, f), for o-ray (a, c, e) and e-ray (b, d, f), respectively. The black and blue lines show the sigma-clipped average and median, respectively (left ordinate) and the  lines show the sigma-clipped standard deviation (right ordinate), for the RoIs (green regions in Fig. \ref{fig:fignicfitsanatomy}). The \texttt{INSROT} on top shows the instrument rotator angle at the exposures.}
  \label{fig:flatchange}
\end{figure*}

For the assumption ($ F^{i j} $ is constant) to hold, one has to select flat frames taken at a similar time, the same HWP angles (\texttt{POL-AGL1}), and the same \texttt{INSROT} because we only used the polarimetric flat for this calculation. The flat flux changes during the exposures (Fig. \ref{fig:flatchange}). The large changes after every 80 frames are due to the change in \texttt{INSROT}, and the scatter within the same \texttt{INSROT} is due to the change in \texttt{POL-AGL1} (semi-periodic for every 4 frames) and/or the true change in the dome flat.

\begin{figure*} [tb!]
  \begin{center}
    \includegraphics[width=0.7\linewidth]{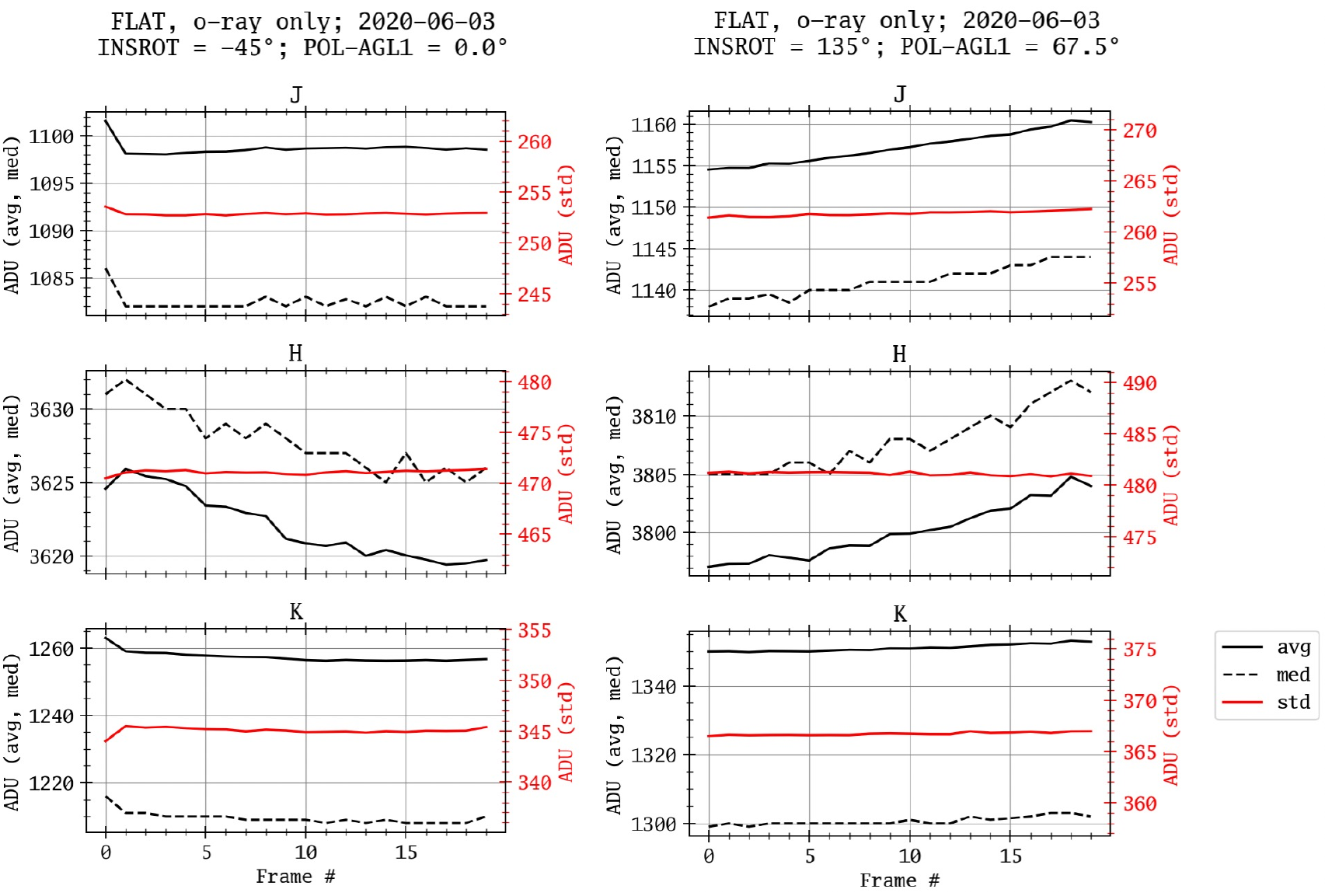}
  \end{center}
  \caption{The change in flat frame pixel statistics after sigma-clipping. The black and blue lines show the average and median, respectively (left ordinate) and the red lines show the standard deviation (right ordinate) for the o-ray RoI (the green region in Fig. \ref{fig:fignicfitsanatomy}).}
  \label{fig:flatchange1}
\end{figure*}

There are 20 consecutive dome flat exposures for the same \texttt{POL-AGL1} and same \texttt{INSROT} for each filter. For a , a few flats with the same \texttt{(POL-AGL1, INSROT)} were selected, and the change in the pixel statistics was calculated. Fig. \ref{fig:flatchange1} shows two such cases. The pixel value changes by $ \lesssim 1\% $ level. Hence, for those 20 frames, it can be assumed that $ F^{i j} $ is constant.

\begin{figure} [tb!]
  \begin{center}
    \includegraphics[width=\linewidth]{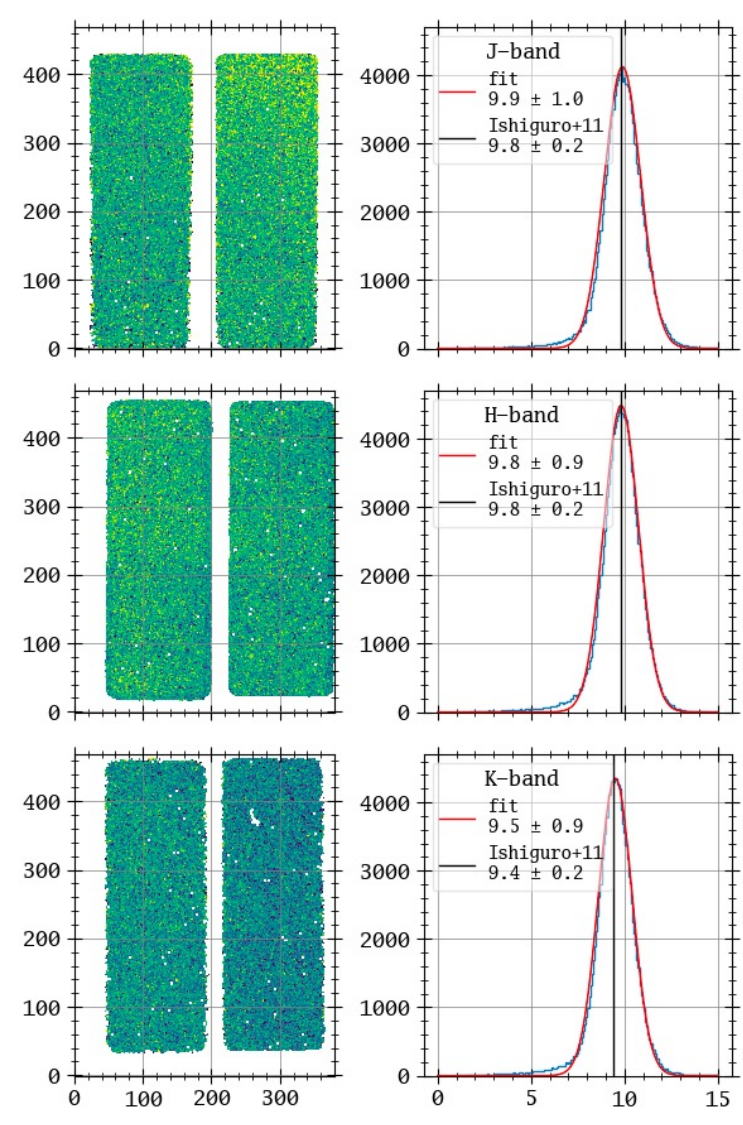}
  \end{center}
  \caption{The 2-D gain maps for J-band (top), H-band (middle), and K-band (bottom). The white pixels are the masked locations (NaN value). The right column shows the corresponding gain factor distribution ([e/ADU]), best-fit Gaussian curve, and the previous result \citep{Ishiguro2011ARNHAO}. In the legend are the best-fit mean and standard deviations.}
  \label{fig:gain}
\end{figure}

The median and standard deviation were obtained for each pixel for 20 frames having the same \texttt{(POL-AGL1, INSROT)}. Then any pixels with too high standard deviation or too low pixel values are removed. Next, the gain factor is calculated using Eq. (\ref{eq: gain calc}) to obtain a 2-D gain map. For all the 32 combinations of \texttt{POL-AGL1} and \texttt{INSROT}, the 2-D map is generated for three detectors. These 32 maps are then median combined with 3-sigma clipping. The final gain map and the gain value distributions are shown in Fig. \ref{fig:gain}.

\subsection{Summary and Comparison with Previous Work}
We have shown that $ R/g $ is independent of location on the detector and very similar for all three detectors of NIC. Whether both $ R $ and $ g $, not only $ R/g $, are indeed constant would be confirmed separately in future work. However, if we assume the two parameters are indeed constant, we obtain gain and readout noise values for all detectors. In Table \ref{tab: gR value}, the gain and readout noise parameters from this work are compared with the previous work \citep{Ishiguro2011ARNHAO}. 

\begin{table*}[tb!]
  \begin{center}
    \begin{tabular}{cc|ccc|c}
      \hline
       & & \multicolumn{2}{c}{\citet{Ishiguro2011ARNHAO}} & This work & Adopted value \\
      Parameter & Detector  & Motor ON & Motor OFF & ($\mathrm{mean} \pm \mathrm{std} $) & in \texttt{NICpolpy}\\
      \hline
                   &J-band & $ 4.7 \pm 0.5 $ & $ 5.4 \pm 0.5 $ &                     & \\ 
      $ R/g $ [ADU]&H-band & $ 8.8 \pm 0.6 $ & $ 7.7 \pm 0.4 $ &  $ 3.7 (\pm <0.1) $ & $ (3.7) $\\
                   &K-band & N/A             & $ 8.8 \pm 0.6 $ &                     & \\
      \hline
                   &J-band & $ 9.8 \pm 0.2 $ & $ 9.2 \pm 0.2 $ & $ 9.9 \pm 1.0 $ & $ 9.9 $\\ 
      $ g $ [e/ADU]&H-band & $ 9.5 \pm 0.2 $ & $ 9.8 \pm 0.2 $ & $ 9.8 \pm 0.9 $ & $ 9.8 $\\
                   &K-band & N/A             & $ 9.4 \pm 0.2 $ & $ 9.5 \pm 0.9 $ & $ 9.5 $\\
      \hline
                   &J-band & $ 46 \pm 5 $ & $ 50 \pm 4 $ & $ 37 \pm 4 $ & $ 37 $ \\ 
      $ R $ [e]    &H-band & $ 84 \pm 5 $ & $ 75 \pm 4 $ & $ 36 \pm 3 $ & $ 36 $ \\
                   &K-band & $ (>250) $   & $ 83 \pm 5 $ & $ 35 \pm 3 $ & $ 35 $ \\
      \hline
      \hline
    \end{tabular}
  \end{center}
  \caption{The parameter values. $ R/g $ in \citet{Ishiguro2011ARNHAO} is calculated from $ R $ and $ g $ with error-propagation, while the $ R $ for this work is calculated from $ R/g $ and $ g $. Error-propagation calculations were done assuming null covariance.}
  \label{tab: gR value}
\end{table*}

In the table, the last column shows the values adopted in \texttt{NICpolpy} (callable as \texttt{nic.GAIN} and \texttt{NIC.RDNOISE}). The official values had been the values in the column ``Motor OFF'', which is the measurement with the shutter motor power turned down (data taken on UT 2011-07-28). 

This discrepancy is likely due to the method used for determining $ R $. They assumed half the variance  the flat difference image is a linear function of signal (Fig. 2 of \cite{Ishiguro2011ARNHAO}),  the Poisson noise regime. The readout noise (or the variance without light) is obtained by extrapolating it to zero signal. This extrapolation, however, works only if the pixels in the flat frame used to calculate variance have nearly identical pixel sensitivity (so that they indeed have the same true flat value). Moreover, in reality, there is always the $ R/g $ term in the variance, so extrapolation using the equation ignoring $ R/g $ may have  in an erroneous estimation of $ R $. However, we note that the more important value, gain (the conversion factor), was determined correctly.

\appendix
\section{Proof of Janesick's Method} \label{s: Janesick}
Consider that  flat and the bias images are $ F $ and $ B $, and any operation (e.g., addition or subtraction) means a pixel-wise operation. Also, any artifacts, bad-pixels or cosmic-rays are ignored. If the true bias level is $ b $ in ADU, any bias image will follow a normal distribution with readout noise:
\begin{equation}
  B \sim \mathcal{N} \left( b, \left( R/g \right)^2 \right) ~\mathrm{[ADU]} ~,
\end{equation}
so that
\begin{equation}
  B_1 - B_2 \sim \mathcal{N} \left ( 0, 2 \left ( R/g \right )^2 \right ) ~\mathrm{[ADU]} ~.
\end{equation}
Hence, 
\begin{equation}
  \sigma^2_{B_1 - B_2} 
    = 2 (R/g)^2 ~,
\end{equation}
so Eq. (\ref{eq: janesick r}) is proven. Since the left-hand side here is approximated by the sample standard deviation, one may write $ \sigma^2_{B_1 - B_2} \approx $ \texttt{np.std(B1-B2, ddof=1)**2} in python.

The (raw) flat image consists of photons with dark plus bias level. Therefore, if $ f $ is the true flat level \emph{plus dark} in ADU, any flat will follow a Poisson distribution, which is roughly identical to a normal distribution when pixel values are high: 
\begin{equation}
  F \sim \mathcal{N} \left ( f + b, \frac{f}{g} + \left (\frac{R}{g} \right )^2 \right ) ~\mathrm{[ADU]}
\end{equation}
which means
\begin{equation}
  F_1 - F_2 \sim 
    \mathcal{N} \left ( 0, 2\frac{f}{g} + 2 \left (\frac{R}{g}\right )^2 \right ) ~,
\end{equation}
and
\begin{equation}
  (F_1 + F_2) - (B_1 + B_2) \sim 
    \mathcal{N} \left (2f, 2 \frac{f}{g} + 4 \left ( \frac{R}{g} \right )^2 \right) ~.
\end{equation}
The $ f/g $ terms are the Poisson noise term, and the $ R/g $ terms are the readout noise term, respectively. From these,
\begin{equation}
  \sigma^2_{F_1 - F_2} - \sigma^2_{B_1 - B_2} 
    \approx 2\frac{f}{g} ~,
\end{equation}
and
\begin{equation}
  (\bar{F}_1 + \bar{F}_2) - (\bar{B}_1 + \bar{B}_2)
   \approx 2f ~.
\end{equation}
This proves Eq (\ref{eq: janesick g}). Since the $ \sigma $ values are approximated by the sample standard deviation, $ \sigma^2_{F_1 - F_2} - \sigma^2_{B_1 - B_2} \approx $ \texttt{np.std(F1-F2, ddof=1)**2 - np.std(B1-B2, ddof=1)**2} and  $ (\bar{F}_1 + \bar{F}_2) - (\bar{B}_1 + \bar{B}_2) = $ \texttt{np.mean((F1+F2)-(B1+B2))} in python.

\section{Supporting Materials} \label{s: sm}
The basic calibration files (master flats for UT 2018-05-07 and 2020-06-04 and \texttt{imask}), flat analyses results (Sect. \ref{s: flat}), and the data reduction example with sample data (Sect. \ref{s: code usage}) are available via the GitHub service\footnote{\texttt{https://github.com/ysBach/NICpolpy\_sag22sm}}. The contents are:
\begin{itemize}
\item \texttt{cal-flat\_20180507-lv1/}: The master flats taken on UT 2018-05-07.
\item \texttt{cal-flat\_20200603-lv1/}: The master flats taken on UT 2020-06-04.
\item \texttt{masks/}: The \texttt{imask} frames (Sect. \ref{s: lv1}).
\item \texttt{flat\_analyses}: The figures and videos of flat maps and their change over \texttt{INSROT} or \texttt{POL-AGL1}. \texttt{sm1.pptx} file here describes the details of how the figures are generated.
\item \texttt{example/}: Sample data, reduction code and output log directory contents (Sect. \ref{s: code usage}).
\end{itemize}

\end{document}